\documentclass[a4paper,11pt]{article}
\usepackage{jheppub} % for details on the use of the package, please see the JINST-author-manual
\usepackage{comment}
\usepackage{physics}
\usepackage{todonotes}
\usepackage{dsfont}
\usepackage{slashed}
\usepackage{hyperref}
\usepackage{url}

\title{The Rise and Fall of the Standard-Model Higgs: \\  Electroweak Vacuum Stability during Kination}

\author[a]{Giorgio Laverda,}
\author[b]{Javier Rubio}

\affiliation[a]{Centro de Astrofísica e Gravitação  - CENTRA,
Departamento de Física, Instituto Superior Técnico - IST,
Universidade de Lisboa - UL, Av. Rovisco Pais 1, 1049-001 Lisboa, Portugal.}
\affiliation[b]{Departamento de Física Teórica and Instituto de Física de Partículas y del Cosmos (IPARCOS-UCM), Universidad Complutense de Madrid, 28040 
Madrid, Spain}

\emailAdd{giorgio.laverda@tecnico.ulisboa.pt}
\emailAdd{javier.rubio@ucm.es}

\abstract{In this paper we investigate the vacuum stability of the non-minimally coupled Standard-Model Higgs during a phase of kinetic domination following the end of inflation. The non-minimal coupling to curvature stabilises the Higgs fluctuations during inflation while driving them towards the instability scale during kination, when they can classically overcome the potential barrier separating the false electroweak vacuum from the true one at super-Planckian field values. Avoiding the instability of the Standard-Model vacuum sets an upper bound on the inflationary scale that depends both on the strength of the non-minimal interaction and on the top quark Yukawa coupling. Classical vacuum stability is guaranteed if the gravitationally-produced energy density is smaller than the height of barrier in the effective potential. Interestingly enough, thanks to the explosive particle production in the tachyonic phase, the Higgs itself can be also appointed to the role of reheaton field responsible for the onset of the hot Big Bang era, setting an additional lower bound on the inflationary scale $\mathcal{H}_{\rm inf} \gtrsim 10^{5.5} \text{ GeV}$. Overall, these constraints favour lower masses for the top quark, in agreement with the current measurements of the top quark pole mass. We perform our analysis semi-analytically in terms of the one-loop and three-loop running of the Standard-Model Higgs self-coupling and make use of lattice-based parametric formulas for studying the (re)heating phase derived in \href{https://arxiv.org/abs/2307.03774}{arXiv:2307.03774}. For a specific choice of $m_t=171.3 \text{ GeV}$ we perform also an extensive numerical scanning of the parameter space via classical lattice simulations, identifying stable/unstable regions and supporting the previous analytical arguments. For this fiducial value, the heating of the Universe is achieved at temperatures in the range $10^{-2} - 10^9 \text{ GeV}$.}

\begin{document}

\maketitle
\flushbottom

%%%%%%%%%%%%%%%%%%%%%%%%%%%%%%%%%%%%%%%%%%%%%%%%%%%%%%%%%%%%%%%%%%%%%%%%%%%
\section{Introduction}
%%%%%%%%%%%%%%%%%%%%%%%%%%%%%%%%%%%%%%%%%%%%%%%%%%%%%%%%%%%%%%%%%%%%%%%%%%%

The Higgs boson occupies a peculiar place in the current paradigm of fundamental physics, sitting at the crossroad of phenomena presently  investigated at particle accelerators and some yet unknown ultraviolet completion of the Standard Model (SM) and gravity. Its first detection in 2012 \cite{ATLAS:2012yve,CMS:2012qbp} led to the confirmation that the electroweak (EW) vacuum we live in might not be absolutely stable and that, at high energy scales, the Higgs field has a chance to undergo a phase transition from the (false) EW vacuum to a true one at super-Planckian field values \cite{Tang:2013bz}. Clearly, the non-observation of such a catastrophic event reinforces the idea that the usual SM vacuum might be stable after all, or at least metastable enough not to decay on a timescale comparable to the age of the Universe. 

The culprit is the problematic running of the renormalised Higgs self-coupling $\lambda(\mu)$, which may become negative already at energy scales of the order of $10^8 \text{ GeV}$. This behaviour depends very sensitively on the top quark Yukawa coupling \cite{Bezrukov:2014ina, Hiller:2024zjp}, which contributes with a large negative correction. 
Several non-trivial effects, ranging from Beyond-the-Standard-Model (BSM) interactions \cite{Elias-Miro:2012eoi,Espinosa:2013lma, Branchina:2013jra, Branchina:2014rva, Branchina:2014usa,Domenech:2020yjf} to higher-order curvature-related operators \cite{Bezrukov:2014ipa,Markkanen:2018pdo, Markkanen:2018bfx}, can be introduced to favour the global convexity of the Higgs effective potential. However, even within the SM alone, a lower value of the top-quark Yukawa coupling is still fully compatible with the absolute stability of the EW vacuum \cite{Bezrukov:2014ipa}. This is perhaps the most crucial piece of information to take into account when analysing the Higgs effective potential at high energies, especially in light of the ${\mathcal{O}}$(GeV) theoretical uncertainties \cite{Bezrukov:2014ina} affecting the relation between the \emph{Monte-Carlo reconstructed top quark mass}  \cite{ATLAS:2018fwq, CMS:2015lbj, CDF:2016vzt, CMS:2018quc, CMS:2023ebf}  obtained using event generators such as PYTHIA or HERWIG \cite{Bierlich:2022pfr,Bellm:2015jjp}
%different definitions and values 
%of the \emph{top quark reconstructed mass} \cite{ATLAS:2018fwq, CMS:2015lbj, CDF:2016vzt, CMS:2018quc, CMS:2023ebf} 
and the \emph{top quark pole mass} actually entering the renormalisation group equations \cite{Myllymaki:2024uje, CMS:2019esx}. Interestingly enough, the latter quantity is found to be consistently lower than the former in the most recent measurements, with $m_t^{\rm pole}=170.5 \pm 0.8 \text{ GeV}$ \cite{CMS:2019esx} and $m_t^{\rm rec}=171.77 \pm 0.37 \text{ GeV}$ \cite{CMS:2023ebf}. This fact automatically renders the EW vacuum more stable and shifts the instability problem to higher energy scales \cite{Steingasser:2023ugv}. 

The magnitude of the instability scale places the vacuum-stability problem within the realm of early-Universe physics, offering a link between the SM parameters and the inflationary and post-inflationary cosmology \cite{Espinosa:2015qea, Strumia:2022kez}. Our aim in this paper is to study the role played by the SM Higgs in a phase of kinetic domination after inflation. To this end, we consider a spectator Higgs field coupled both to the ordinary SM content and to the Ricci scalar via a non-minimal gravitational interaction. 
As shown in Refs.~\cite{Figueroa:2016dsc, Nakama:2018gll, Dimopoulos:2018wfg, Bettoni:2018utf, Bettoni:2018pbl}, this specific setting gives rise to a Hubble-induced symmetry breaking soon after the end of inflation and to the associated displacement of the spectator field to large field values, significantly amplifying its fluctuations in a rapid burst that lasts less than one $e$-fold \cite{Bettoni:2019dcw,Bettoni:2021zhq,Laverda:2023uqv,Bettoni:2021qfs}.~\footnote{Note that this dynamics is substantially different from the familiar picture of a resonantly-excited Higgs field coupled to an oscillating background \cite{Figueroa:2015rqa, Figueroa:2016dsc, Figueroa:2017slm, Figueroa:2016wxr, Herranen:2015ima, Mantziris:2020rzh, Mantziris:2021oah, Mantziris:2021zox, Mantziris:2022bfe, Mantziris:2022fuu, Mantziris:2023xsp, Kohri:2016wof, Postma:2017hbk, Ema:2017loe, Enqvist:2016mqj, Ema:2016kpf}, where the amplification is less efficient and typically lasts for a much longer period.} Only at the moment of peak tachyonic production the problem of vacuum stability enters the game. In particular, the rising energy-density of the Higgs fluctuations has to be checked against the height of the barrier separating the false and the true vacuum, with the possibility of constraining the parameter space of the theory. If the Higgs vacuum is classically stable at this precise moment, it will stay so for the following cosmological history. On the other hand, the patches that classically fall toward the global minimum cannot be rescued, unless some additional mechanism restores the convexity of the effective potential at large field values \cite{Bezrukov:2014bra, Espinosa:2015qea, Bezrukov:2014ipa}.

The present work focuses on some fundamental questions. Can the non-minimally-coupled spectator Higgs field be stable during kination? If so, can it be also responsible for heating the Universe after inflation? What does this identification imply for the model parameters? We perform a semi-analytical analysis using the one-loop and three-loop renormalisation-group (RG) running of the Higgs self-coupling and their approximations around the instability scale. Because of the significant dependence of the running on the top quark mass, we maintain an agnostic approach that encompasses top quark masses ranging between $170-173 \text{ GeV}$. The lattice-based parametric formulas derived in \cite{Laverda:2023uqv} give us a convenient way to estimate the probability that the tachyonically-enhanced Higgs crosses the barrier in the effective potential. The same set of parametric formulas can also be used to enforce the successful achievement of (re)heating before Big Bang Nucleosynthesis (BBN). The stable and unstable regions of the parameter space are extensively scanned by more than one thousand 3+1-dimensional classical lattice simulations, using the publicly-available \texttt{$\mathcal{C}osmo\mathcal{L}attice$} code \cite{Figueroa:2020rrl, Figueroa:2021yhd}. Both the analytical approach and the numerical one indicate that a low top quark mass is necessary for the successful reheating of the post-inflationary Universe while preserving the stability of the Higgs vacuum. 

The paper is organised as follows. After some preliminary description of the setup in Section \ref{sec:setup}, we discuss the vacuum stability problem in Section \ref{sec:stability_1loop} focusing on the one-loop running of the Higgs self-coupling. We also analyse the constraints coming from quantum metastability and from post-inflationary (re)heating, the latter in Section \ref{sec:reheating_constraints}. Section
\ref{sec:stability_3loop} contains the study of vacuum stability at three-loops, which leads to setting constraints on the top quark mass at the EW scale. We summarise our findings in Section \ref{sec:conclusions}. Appendix \ref{app:higgs_coupled_sm} contains more details about perturbative and non-perturbative Higgs decay channels in the early Universe, while Appendix \ref{app:lattice} describes the setup behind our lattice simulations.   

%%%%%%%%%%%%%%%%%%%%%%%%%%%%%%%%%%%%%%%%%%%%%%%%%%%%%%%%%%%%%%%%%%%%%%%%%%%
\section{An alternative Higgstory} \label{sec:setup}
%%%%%%%%%%%%%%%%%%%%%%%%%%%%%%%%%%%%%%%%%%%%%%%%%%%%%%%%%%%%%%%%%%%%%%%%%%%

The question whether the EW vacuum is stable or not offers a chance to link together the fundamental couplings in the SM and the parameters of early-Universe cosmology. We investigate this issue starting from a minimal version of the SM with no new physics between the EW scale and the Planck scale. The Higgs doublet $H$ is therefore the only scalar degree of freedom besides the inflaton field on the Universe's gravitational background. A general feature of any renormalised energy-momentum tensor of a quantum field in curved spacetime is the appearance of curvature-dependent counter terms \cite{Birrell:1982ix, Mukhanov:2007zz}. Since these terms cannot be set to vanish at all scales, we will consider a non-minimal coupling between the Higgs and the scalar curvature $\sim \xi R H^{\dagger}H$. The resulting model Lagrangian,
\begin{equation} \label{eq:higgs_lagrangian}
    \frac{\mathcal{L}}{\sqrt{-g}}=  \frac{M^2_{\rm P}}{2} R - g^{\mu \nu} (D_{\mu}H)^{\dagger}(D_{\nu}H) - \lambda \left( H^{\dagger}H - \frac{v_{\rm EW}^2}{2} \right)^2 - \xi H^{\dagger}H R  + \mathcal{L}_{\rm SM} + \mathcal{L}_{\phi}\,,
\end{equation}
contains also the SM one without the Higgs sector, ${\cal L}_{\rm SM}$, and an inflationary sector ${\cal L}_{\phi}$, with $M_P=2.4\times 10^{18}$ GeV the reduced Planck mass and $v_{\rm EW}$ the electroweak Higgs expectation value. We assume the Higgs field to be energetically subdominant with respect to the inflaton at the energy scales under consideration, i.e. the Higgs plays the role of a spectator scalar field directly coupled to the background geometry \cite{Bettoni:2019dcw, Bettoni:2021zhq, Laverda:2023uqv, Dimopoulos:2018wfg, Nakama:2018gll} and its contribution to the overall Ricci-scalar prefactor is set to be subdominant as compared to the usual Planck mass counterpart, namely $\xi H^{\dagger}H \ll M^2_P$. Given this energetic subdominance, we do not need to spell out a specific inflationary theory  ${\cal L}_{\phi}$, being sufficient to consider a global cosmological equation of state dictating the rate of expansion. 

The changes in the scalar curvature $R$ throughout inflation and (pre)heating induce the time-dependence of the Higgs effective mass, which can result in an enhancement or quenching of its quantum modes. In the present scenario,  we assume that the inflaton energy density is given primarily by its kinetic component, which leads to a so-called \emph{kination} epoch, when the Universe expands with a stiff equation of state ($w_{\rm kin}=1$).
This phase is a common occurrence of various inflationary models, especially those exhibiting a non-oscillatory character,\footnote{For a discussion on kination arising in a string-cosmology setting, see \cite{Apers:2022cyl, Apers:2024ffe, Revello:2023hro, Conlon:2022pnx}.} such as \emph{Quintessential Inflation} \cite{Wetterich:1987fm,Wetterich:1994bg,Peebles:1998qn,Spokoiny:1993kt,Brax:2005uf,BuenoSanchez:2007jxm,Wetterich:2013jsa,Wetterich:2014gaa, Hossain:2014xha,Agarwal:2017wxo,Geng:2017mic,Dimopoulos:2017zvq,Rubio:2017gty,Dimopoulos:2017tud,Akrami:2017cir,Garcia-Garcia:2018hlc}  (see \cite{Bettoni:2021qfs} for a review), where this stiff expansion can last for several $e$-folds. In these settings, the quintessential inflationary potential interpolates between two plateaus, a high-scale one to inflate the early Universe and a low-scale one to explain the cosmological constant in the late Universe. A kinetic-energy-dominated phase takes place between the two, as most of the inflaton potential energy is converted into kinetic energy. For the sake of simplicity, we assume the transition between inflation and kination to be instantaneous or, at least, much faster than the typical duration of the heating processes that follow it. Together with the aforementioned energetic subdominance, this approximation allows us to study the dynamics of the Higgs field independently of the specific shape of the inflationary potential.

The non-minimal interaction with gravity and the time-dependence of the Ricci scalar define two distinct dynamical phases. Throughout the inflationary stage, the Higgs field is stabilised against the amplification of quantum fluctuations \cite{Herranen:2014cua, Gialamas:2022gxv, Gialamas:2023emn, Kohri:2016wof}. The Ricci scalar in an expanding FLRW background,
\begin{equation}\label{eq:ricciscalar}
    R=6\left(\dot{\mathcal{H}} + 2\mathcal{H}^2 \right) = 3(1-3w_{\phi})\mathcal{H}^2\,, 
\end{equation}
contributes indeed as a large mass term, since $R=12\mathcal{H}^2$ during inflation ($w_{\phi}=-1$), with $\mathcal{H}=\dot{a}/a$ denoting the usual Hubble function. Estimating the amplitude of the Higgs inflationary fluctuations shows us that the average amplitude of the field computed from the super-horizon power spectrum depends on the inverse of the non-minimal coupling parameter \cite{Opferkuch:2019zbd, Cosme:2018nly, Riotto:2002yw, Kohri:2016wof}, namely
\begin{equation}
    \langle h_{\rm inf}^2 \rangle \simeq \frac{1}{\sqrt{3\xi}}\frac{\mathcal{H}_{\rm inf}^2}{24\pi^2} \; ,
    \label{eq:inf_flauctuation_amplitude}
\end{equation}
with $\mathcal{H}_{\rm inf}$ the inflationary scale and $h$ the radial component of the Higgs field in the unitary gauge $H=(0,\,h/\sqrt{2})^T$. Neither these super-horizon fluctuations nor the much-smaller, exponentially-suppressed sub-horizon fluctuations \cite{Ford:2021syk} can probe the instability region for $\xi \gg 1$ and $\mathcal{H}_{\rm inf}$ below the vacuum instability scale. Moreover, the other scale in the problem, i.e. the field transition value from a quadratic to a quartic potential, $h\sim \sqrt{\xi R / \lambda}$, is orders-of-magnitude larger than the amplitude in \eqref{eq:inf_flauctuation_amplitude}, which can be taken as a conservative estimate. The Higgs fluctuations \eqref{eq:inf_flauctuation_amplitude} amount also to small isocurvature perturbations for $\xi \gtrsim 0.1$ \cite{Bettoni:2021zhq, Bettoni:2018utf, Opferkuch:2019zbd} with a negligible effect on the adiabatic metric fluctuations. Therefore, we safely assume the Higgs to be a homogeneous field approximately confined to its vacuum at the end of inflation \cite{Felder:2001kt}.~\footnote{More precisely, the initial conditions at the onset of kination depend on the dynamics of the transition. In particular, embedding our setup into a specific inflationary potential would associate a speed to the inflation-kination transition, leading potentially to a transient non-adiabatic amplification of Higgs modes \cite{Ford:2021syk, Kolb:2023ydq}. However, the associated particle production during the transition turns out to be marginal as compared to the symmetry-breaking mechanism considered in this paper \cite{Bettoni:2021zhq, Bettoni:2019dcw}. As we will see, the tachyonic amplification is not only much more efficient but also more long-lasting than the brief violation of adiabaticity taking place during the transition. }

\begin{figure}[tb]
\centering
\includegraphics[width=.8\textwidth]{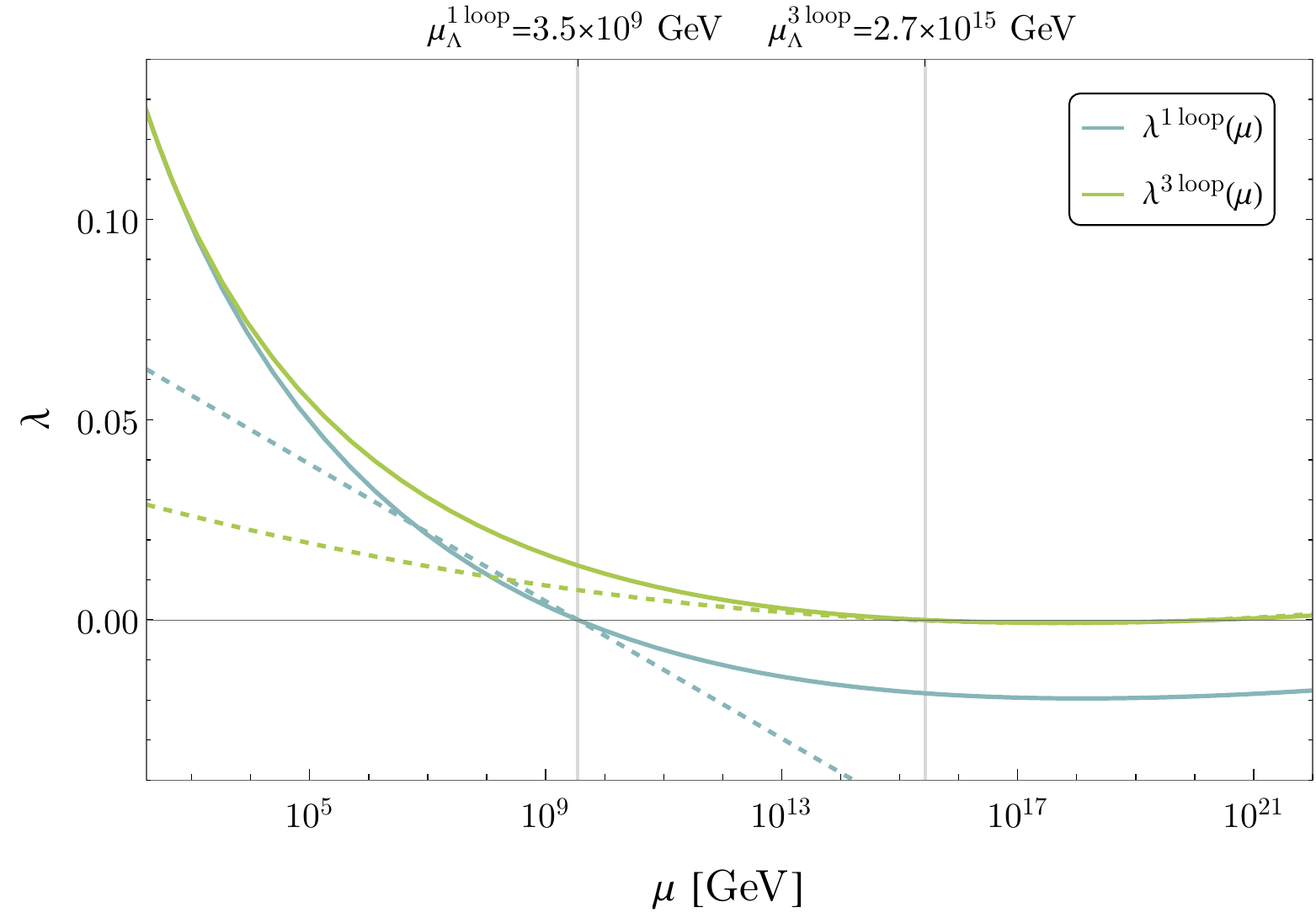}
\caption{Running of the SM Higgs self-coupling at one (blue solid line) and three loops (green solid line), assuming EW values $m_h=125.5 \text{ GeV}$ and $m_t=171.3 \text{ GeV}$ for the Higgs and the top quark mass, respectively. The corresponding dashed lines approximate the running around the instability scale as $\sim \ln \mu$ in the one-loop case and as $\sim \ln^2\mu$ in the three-loop case. Both running have been computed with the Mathematica package in \cite{BezrukovNotebook}.} \label{fig:lambda_running}
\end{figure}

The situation in kination is dramatically different. If during inflation the non-minimal coupling has a stabilising effect, during kination it can catastrophically destabilise the Higgs potential. For a stiff equation of state parameter $w_{\phi}=1$, the Ricci scalar becomes negative ($R=-6\mathcal{H}^2$) and contributes as a large tachyonic mass term for $\xi \gg 1$. Local minima of the potential develop then at $h\sim \sqrt{-\xi R / \lambda}$. The value of the scale-dependent renormalised Higgs self-coupling $\lambda$ in this expression is obtained via the RG equations at different loop expansions. In particular, we focus on the one-loop and three-loop beta functions in Figure~\ref{fig:lambda_running}, as they both admit simple approximations around the instability scale. Both runnings depend very sensitively on the mass of the top quark $m_t$. However, once its value is specified, the instability scale $\mu_{\Lambda}$ becomes uniquely identified.\footnote{In the remainder of this work, we assume a fixed value of $m_h=125.5 \text{ GeV}$ and we set $m_t=171.3 \text{ GeV}$ when a specific top mass has to be chosen. This value of the top quark mass lays in the upper limit of the confidence interval in \cite{CMS:2019esx} and approaches the lower limit of the measurement in \cite{CMS:2023ebf}. } Comparing the instability scale to the position of the local minima reveals two different scenarios. If the minima are displaced at sufficiently large values of the field, the typical potential barrier at $\mu_{\Lambda}$ disappears and the Higgs is allowed to reach arbitrarily large amplitudes. However, if the local minima exist at scales lower than $\mu_{\Lambda}$, the tachyonic instability amplifies the Higgs fluctuations, which can classically overcome the barrier. In this regime, the issue of vacuum stability has to be investigated by searching the parameter space for the stable configurations. 

The tachyonically-enhanced Higgs energy-density evolves as radiation $\sim a^{-4}$ over the decaying inflaton background energy $\sim a^{-6}$, eventually overtaking it \cite{Allahverdi:2020bys,Bettoni:2021qfs}. If stability is enforced, this natural mechanism can lead to the onset of radiation domination within a few $e$-folds of post-inflationary expansion \cite{Bettoni:2021zhq}. In other words, the higher degree of instability caused by the non-minimal coupling to curvature has the advantage of producing a sufficient number of particles to heat the Universe in the Higgs sector alone, i.e. without invoking unknown inflaton couplings or introducing additional degrees of freedom beyond SM. Identifying the Higgs with a so-called \textit{reheaton} field imposes extra constraints on the scale of kination $\mathcal{H}_{\rm kin}$ and, given that in our setup $\mathcal{H}_{\rm kin} \approx \mathcal{H}_{\rm inf}$, we obtain a lower bound on the inflationary scale as well. \footnote{Approximating $\mathcal{H}_{\rm inf}$ to be constant allows us to derive constraints independently of the exact inflationary model. However, the stability of the EW vacuum can be influenced by a departure from perfect de Sitter expansion caused by Planck-suppressed operators, see \cite{Fumagalli:2019ohr} for a detailed analysis.} The non-minimal coupling term becomes secondary over time as it decreases with $\mathcal{H}^2$, the local minima approach the origin and the Higgs begins its oscillatory phase in an almost-quartic potential. After about one $e$-fold of tachyonic instability, parametric resonance can lead to the production of SM particles, especially of gauge bosons \cite{Enqvist:2013kaa, Enqvist:2016mqj}, since the production of fermions is suppressed by Pauli blocking effects \cite{Greene:2000ew}. This process can last several $e$-folds and populates the Universe with SM particles before the end of the heating phase. In spite of this, the macroscopic dynamics of the heating stage is due almost entirely to the Higgs, as we show in Appendix \ref{app:higgs_coupled_sm}. The typical energy-density of the daughter particles produced within the first oscillations of the Higgs is only a small fraction of the Higgs energy-density and, therefore, does not play a significant role in the heating dynamics. Nonetheless, resonant production enhances the daughter fields energy-density over time, together with a purely thermodynamical redistribution of energy within the primordial plasma, which lasts $\mathcal{O}(10)$ $e$-folds \cite{Bettoni:2021zhq,Laverda:2023uqv}. 

%%%%%%%%%%%%%%%%%%%%%%%%%%%%%%%%%%%%%%%%%%%%%%%%%%%%%%%%%%%%%%%%%%%%%%%%%%%
\section{Vacuum stability at one loop} \label{sec:stability_1loop}
%%%%%%%%%%%%%%%%%%%%%%%%%%%%%%%%%%%%%%%%%%%%%%%%%%%%%%%%%%%%%%%%%%%%%%%%%%%

In this section, we pick up the discussion on the stability of the Higgs vacuum during kination by studying the dynamics of the Higgs in its one-loop renormalisation-group-improved (RGI) potential. This setup comes with the major benefit of allowing an analytical understanding of several key constraints. The starting point of our study is the Lagrangian in \eqref{eq:higgs_lagrangian}, which contains the usual potential for the SM Higgs in the unitary gauge at scales much above the EW scale $h \gg v_{\rm EW} = 246 \text{ GeV}$,
\begin{equation}\label{eq:effective_potential}
    V(h) \simeq \frac{\lambda}{4}h^4 \;.
\end{equation}
The well-known renormalisation procedure of the SM sector leads to the energy-scale dependence of the observable couplings. The RG running of the Higgs self-coupling $\lambda(\mu)$ at one loop is given by \cite{Bezrukov:2014ina}
\begin{equation} \label{eq:lambda_running_1loop}
    \beta_{\lambda}^{\rm 1 loop}=\frac{\partial \lambda(\mu)}{\partial \ln \mu} = \frac{1}{16\pi^2} \left( 24\lambda^2(\mu) + 12y_t^2(\mu)\lambda(\mu) - 6y_t^4(\mu) \right) \; ,
\end{equation}
where we have considered only the Yukawa coupling to the top quark $y_t(\mu)$, as this is the major player in the vacuum instability problem.~\footnote{Besides considering only the top quark, we also neglect potential threshold effects associated to the non-renormalisable character of the Standard Model non-minimally coupled to gravity \cite{Bezrukov:2014ipa,Rubio:2018ogq}, as well as higher curvature corrections only relevant for small values of the non-minimal coupling parameter $\xi\lesssim1$ \cite{Figueroa:2017slm, Markkanen:2018pdo}, far below the range of parameters we consider.}  The solution of \eqref{eq:lambda_running_1loop} requires the knowledge of the RG equation for $y_t(\mu)$. However, as a reasonable approximation around the instability scale $\lambda(\mu_{\Lambda})=0$, we can write \cite{Espinosa:2015qea}
\begin{equation} \label{eq:lambda_running_1loop_apx}
    \frac{\partial \lambda(\mu)}{\partial \ln \mu} \simeq -\frac{6y^4_{\Lambda}}{16\pi^2} \; ,
\end{equation}
where we have set $y_t(\mu) \approx y_{\Lambda}$ to be constant in virtue of its logarithmic running. The running of $\lambda$ around the instability scale is then given by \cite{Bezrukov:2012sa}
\begin{equation}\label{eq:lambda_running_approx}
    \lambda^{\rm 1 loop}(\mu)=\lambda(\mu_0) - \frac{3}{16\pi^2}y^4_{\Lambda} \ln \left( \frac{\mu^2}{\mu_0^2}\right) \simeq - \frac{3}{16\pi^2}y^4_{\Lambda} \ln \left( \frac{h^2}{\mu_{\Lambda}^2}\right)\,,
\end{equation}
with the renormalisation scale $\mu$ chosen as $\mu^2=\mathcal{H}^2 + h^2 \approx h^2$, as customary for models including curvature-dependent interactions \cite{Markkanen:2018bfx, Markkanen:2018pdo}.~\footnote{We expect from \cite{Laverda:2023uqv} that $h^2$ is at least one or two orders of magnitude greater than the Hubble scale after the tachyonic amplification. Since we are interested in the dynamics of the tachyonically-amplified Higgs in the proximity of the the barrier, we set $h^2\gg\mathcal{H}^2$. For the range of $\cal H$ considered in this paper, this approximation remains reliable in a large range $-3\lesssim \log (h/\mu_{\Lambda})\lesssim 0$ around the instability scale. }  The Higgs effective potential receives therefore a logarithmic correction
\begin{equation} \label{eq:effective_potential_approx}
    V_{\rm eff}^{\rm 1 loop} \simeq  \frac12 \xi R h^2 - \frac{3}{64\pi^2}y^4_{\Lambda}h^4 \ln \left( \frac{h^2}{\mu_{\Lambda}^2}\right) \;.
\end{equation}
The approximated logarithmic running of $\lambda(\mu)$ constitutes the pivotal result to constrain the parameter space of the model. In particular, a first constraint arises when considering the appearance of a barrier in the effective potential, which depends on the kination scale $\mathcal{H}_{\rm kin}$ and on the non-minimal coupling parameter $\xi$. A second constraint is derived by requiring the non-crossing of the barrier.

From the running in Figure~\ref{fig:lambda_running}, we conclude that at scales ${\mu_{\Lambda} \approx 10^{8} - 10^{15} \text{ GeV}}$ a barrier can form in the effective potential only if the non-minimal interaction with the curvature scalar is small enough, so that a local minimum exists at scales lower than the instability scale. If the local minimum $h^2_{\rm min}=\xi \mathcal{H}^2/\lambda$ lays in the proximity of the instability scale, $\lambda$ acquires a value that is well-approximated by the expression \eqref{eq:lambda_running_approx}, allowing us to easily find the local maxima and minima of the potential \eqref{eq:effective_potential_approx} by solving $\partial V_{\rm eff}^{\rm 1 loop} / \partial h=0$. Setting $R=-6\mathcal{H}^2$ during kination, we find solutions only when the condition
\begin{equation}\label{eq:barrier_existence}
    \xi < \frac{y_{\Lambda}^4 \mu_{\Lambda}^2 }{32 \, e^{3/2} \pi^2 \, \mathcal{H}^2}
\end{equation}
is fulfilled, i.e. when the local minimum is placed at scales lower than the instability scale. The inequality in \eqref{eq:barrier_existence} sets a first upper bound on the model parameters. Larger values of the non-minimal coupling are ruled out, since the potential acquires everywhere a negative curvature. A quick estimate with some typical values of the parameters ($y_{\Lambda}\approx 0.5$) reveals that $\xi \lesssim 10^{-4}\mu_{\Lambda}^2/\mathcal{H}^2$, i.e. the Hubble scale when the field approaches the barrier has to be at least two orders of magnitude smaller than the instability scale to allow the barrier to be present for $\xi>1$. Within the domain set by \eqref{eq:barrier_existence}, the full solutions are given by
\begin{align}
    h_{\rm max}^2(\xi, y_{\Lambda}, \mathcal{H}, \mu_{\Lambda}) &= \mu_{\Lambda}^2 \exp \left[  W\left(-\frac{32 \sqrt{e} \pi ^2 \, \mathcal{H}^2 \xi}{y_{\Lambda}^4\mu_{\Lambda}^2}\right)-\frac{1}{2}\right] \; , \\
    h_{\rm min}^2(\xi, y_{\Lambda}, \mathcal{H}, \mu_{\Lambda}) &= \mu_{\Lambda}^2 \exp \left[ W_{-1}\left(-\frac{32 \sqrt{e} \pi ^2 \, \mathcal{H} ^2 \xi}{y_{\Lambda}^4\mu_{\Lambda}^2}\right)-\frac{1}{2} \right] \,,
\end{align}
with $W$ the Lambert function. The first two solutions correspond to the local maxima at the top of the barrier, while the last two solutions correspond to the local minima before the barrier.

The second constraint on the stability of the Higgs vacuum is given by the non-crossing of the barrier in the effective potential. As the tachyonic phase enhances the Higgs energy-density, it becomes possible for regions of the Universe to overcome such barrier. The macroscopic evolution can be studied in a classical sense, by comparing the average peak energy produced during the tachyonic phase with the height of the potential barrier. In other words, we check the stability condition
\begin{equation} \label{eq:barrier_overcoming}
   \rho_{\text{tac}}(\lambda(\mu), \xi) < V_{\rm eff}^{\rm 1 loop}(h_{\rm max}(\xi, y_{\Lambda}, \mathcal{H}, \mu_{\Lambda})) 
\end{equation}
for those combinations of parameters that satisfy \eqref{eq:barrier_existence}. The results in \cite{Laverda:2023uqv}, derived from fitting hundreds of fully-fledged classical lattice simulations to a few simple parametric formulas, allow specifying the maximum energy-density produced in the tachyonic phase as a function of $\xi$ and $\lambda(\mu)$, namely
\begin{equation}
    \rho_{\text{tac}}(\lambda(\mu), \xi) = 16 \, \mathcal{H}^4_{\rm kin}\, \exp\left(\beta_1 + \beta_2 \,\nu+ {\beta_3} \ln \nu \right)  \,,
\label{eq:fit_rho_br}
\end{equation} 
with
\begin{equation}
\nu=\sqrt{\frac{3\xi}{2}}\,,    
\end{equation}
and coefficients
\begin{equation}
    \beta_1 = -7.03 - 0.56 \, n  \,,\hspace{7mm}
    \beta_2 = -0.06 - 0.04 \, n \,, \hspace{7mm}
    \beta_3 = 7.15 + 1.10 \, n \,, \label{coeff_rho_br}
\end{equation}
which depend on the Higgs self-coupling via $n=-\log \lambda(\mu)$. It is interesting to notice the hierarchy in the stability conditions, namely that \eqref{eq:barrier_existence} (existence of the barrier) is always fulfilled when \eqref{eq:barrier_overcoming} (overcoming of the barrier) is fulfilled. Indeed, \eqref{eq:barrier_existence} embodies the demand that the non-minimal coupling term in \eqref{eq:effective_potential_approx} is smaller than the quartic term around the instability scale. For a fixed representative value $\Bar{\lambda}$ this means $\xi \lesssim \Bar{\lambda} \mu_{\Lambda}^2 / \mathcal{H}^2 $. On the other hand, requiring the tachyonically-amplified Higgs fluctuations not to approach the instability scales leads to $\xi \lesssim \Bar{\lambda} \mu_{\Lambda}^4 /( \mathcal{H}^2 h^2_{\rm tac})$. From the parametric formulas in \cite{Laverda:2023uqv} we can estimate the amplitude of the Higgs field at the end of the tachyonic phase as
\begin{equation}
    \langle h^2_{\rm tac} (\lambda(\mu), \xi)\rangle = 4 \mathcal{H}^2_{\rm kin} \, \exp\left(\alpha_1 + \alpha_2 \nu +{\alpha_3} \ln \nu \right)\, ,
\end{equation}
with coefficients
\begin{equation}
    \alpha_1 = -4.92 + 0.74 \, n  \,,\hspace{7mm}
    \alpha_2 = -0.04 - 0.02 \, n \,, \hspace{7mm}
    \alpha_3 = 3.54 + 0.61 \, n \,.
\end{equation}
For typical values of $\lambda \sim 10^{-3}-10^{-4}$, $\langle h^2_{\rm tac} \rangle$ is at least four orders of magnitude larger than $\mathcal{H}_{\rm kin}^2$, therefore confirming that the criterion on the not-overcoming of the barrier \eqref{eq:barrier_overcoming} is stricter than \eqref{eq:barrier_existence}. The difference between the two estimates is entirely due to the explosive process of particle production taking place during the tachyonic phase.

\begin{figure}[tb]
\centering
\includegraphics[width=.65\textwidth]{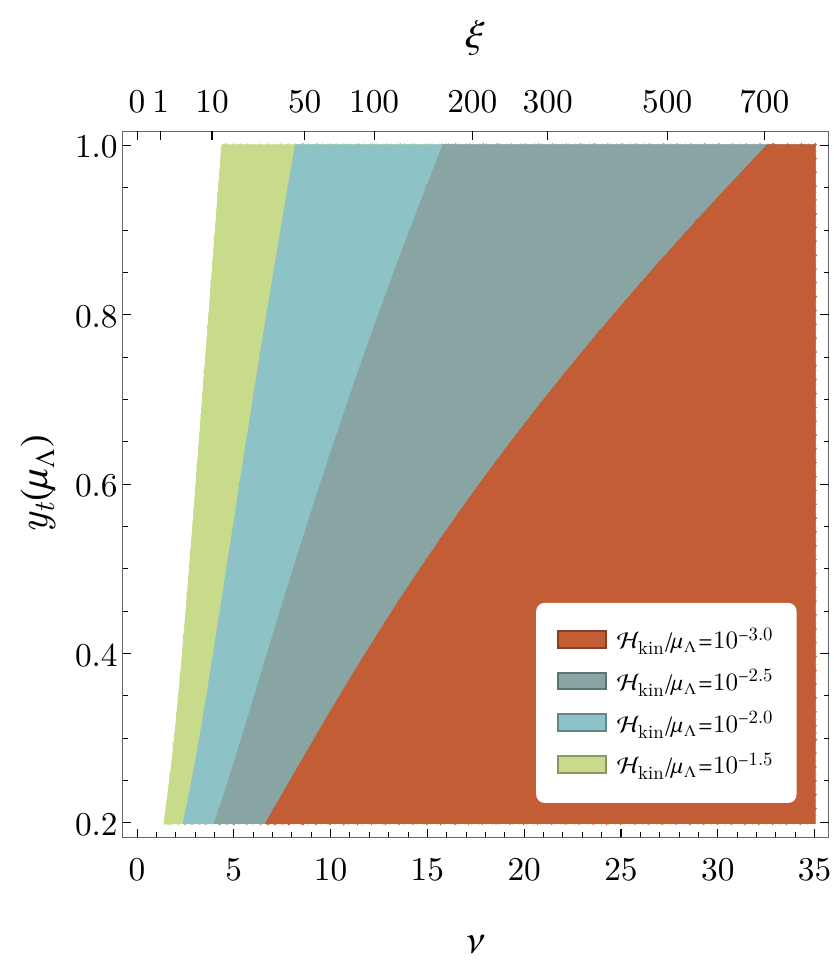}
\caption{Constraints on the top Yukawa coupling at the instability scale $y_t(\mu_{\lambda})$ given by the stability requirement as a function of $\nu=\sqrt{3\xi/2}$. The coloured regions are excluded by the criteria in \eqref{eq:barrier_overcoming} (no overtaking of the barrier). Different colours correspond to different choices of the ratio $\mathcal{H}_{\rm kin}/\mu_{\Lambda}$. Each coloured region includes all the regions to its right. The running of the Higgs self-coupling is given by the approximated one-loop expression in \eqref{eq:lambda_running_approx}.} \label{fig:barrier_exixtence}
\end{figure}

A good approximation for the value of $\lambda(\mu)$ at the end of the tachyonic phase can be obtained by computing
\begin{equation} \label{eq:lambda_min}
    \lambda_{\rm min}^{\rm 1loop}(\xi, y_{\Lambda}, \mathcal{H}, \mu_{\Lambda}) = -\frac{3}{16\pi^2}y_{\Lambda}^4\ln \left[\frac{h_{\rm min}^2(\xi, y_{\Lambda}, \mathcal{H}, \mu_{\Lambda})}{\mu^2_{\Lambda}} \right]\,,
\end{equation}
which corresponds to the magnitude of the Higgs self-coupling at the scale of the local minimum $|h_{\rm min}(\xi, y_{\Lambda}, \mathcal{H}, \mu_{\Lambda})|$. Through the insertion of \eqref{eq:lambda_min} in the parametric formula \eqref{eq:fit_rho_br} we can probe the parameter space of the model. Our findings are summarised in Figure~\ref{fig:barrier_exixtence}, where we display the regions of parameter space that allow for the classical stability of the Higgs field during kination. We focus on a wide range for the non-minimal coupling parameter $\xi \gg 1$, thus considerably extending the analysis in \cite{Figueroa:2016dsc, Figueroa:2017slm} and covering the gap up to Higgs-Inflation-like values \cite{Bezrukov:2007ep,Rubio:2018ogq}.~\footnote{It is important to notice, however, that, beyond the relative sign of the non-minimal coupling, the initial conditions for the Higgs field are strikingly different in these two scenarios. While the onset of Higgs inflation requires Planckian displacements of the field, the current scenario assumes this value to be initially close to zero within the corresponding inflating patch.} Different colours indicate the excluded values of the top Yukawa coupling at the instability scale for different values of ${\cal H}_{\rm kin}/\mu_\Lambda$. Note that the constraints on the value of the Yukawa coupling are set at the instability scale, where $\mu^2_{\Lambda}\simeq h^2$ depends on the non-minimal coupling parameter and on the scale of kination $\mathcal{H}_{\rm kin}$. As a consistency condition of the parametric formulas in \cite{Laverda:2023uqv}, we must impose $\nu \geq 5$ in order to avoid the presence of horizon-reentering modes in the numerical lattice simulations \cite{Bettoni:2019dcw,Bettoni:2021zhq}. In principle, nothing forbids us to chose lower values of $\nu$ in the general framework of this model and we include the region $0 \leq \nu \leq 5$ in our figures as well, bearing in mind that the parametric formulas we are using might not be an accurate description of the physics happening in that portion of the parameter space. We also note here, that all the parameter space in Figure~\ref{fig:barrier_exixtence} satisfies the conditions on energetic subdominance of the Higgs field,
\begin{equation} \label{eq:sub_planck}
\rho_{\rm tac}\ll \rho_\phi\,,\hspace{20mm}     \xi h_{\rm tac}^2 < 0.1 \times M_P^2 \; ,
\end{equation}
where we have set an upper bound on the correction to the Planck mass at $10\%$ level. These conditions represent a self-consistency check within the model, as they quantify the statement that the Higgs has to be a subdominant spectator field before the completion of the heating phase. Because of the typical instability scales $\mu^{\rm 1loop}_{\Lambda}\simeq10^8 - 10^{9} \text{ GeV}$, the constraint in \eqref{eq:barrier_overcoming} is always stricter than those in \eqref{eq:sub_planck}.

On top of the classical fluctuations overcoming the barrier in the potential, quantum tunnelling can further destabilise the Higgs vacuum during the heating phase. Taking into account such effect in the dynamics of the Higgs is not an easy task because of the interplay between classical stochastic fluctuations and the spacetime-dependent quantum decay rate \cite{Chauhan:2023pur}. As a first estimate of the quantum effects, we simply compute the decay rates $\Gamma_{\rm HM}$ and $\Gamma_{\rm CL}$ associated with the Hawking-Moss and Coleman-de Luccia solutions to the bounce equation \cite{Markkanen:2018pdo} and compute the nucleation probability of a single true-vacuum bubble in the heating phase. \footnote{For a study of first order phase transitions during kination, see for instance \cite{Kierkla:2023uzo}.} We assume a constant Hubble rate $\mathcal{H}$ throughout the tunnelling process. Following the discussion in \cite{Markkanen:2018pdo, Markkanen:2018bfx}, the decay rate can be computed as
\begin{equation}
    \Gamma(\mathcal{H}) \sim \mathcal{H}^4 e^{-B(\mathcal{H})}\,, 
\end{equation}
with 
\begin{align}
    B_{\rm HM}(\mathcal{H}) &= \frac{8\pi^2V_{\rm eff}^{\rm 1 loop}(h_{\rm max})}{3\mathcal{H}^4} \,, \\
    B_{\rm CL}(\mathcal{H}) &= \frac{8\pi^2}{3|\lambda(\mu_{\rm min})|} \left[ 1+ 36\left(\xi-\frac16 \right) \frac{\mathcal{H}^2}{\mu_{\rm min}^2}\ln \left( \frac{\mu_{\rm min}}{\mathcal{H}}\right)\right] \; ,
\end{align}
the exponents associated to the Hawking-Moss \cite{Hawking:1981fz} and Coleman-de Luccia instanton solutions \cite{Coleman:1980aw}, $V_{\rm eff}^{\rm 1 loop}(h_{\rm max})$  the height of the barrier in the effective potential and $\mu_{\rm min}$ the energy scale corresponding to the smallest value of $\lambda(\mu)$. Given the typical scales in the analysis, we expect $V_{\rm eff}^{\rm 1 loop}(h_{\rm max})/\mathcal{H}^4 \gg 10^4$ and $\mu_{\rm min}/\mathcal{H}\gtrsim10^9$. The number of bubbles that form during the $N$ $e$-folds of heating has to be smaller than one in order to guarantee quantum metastability in that phase, i.e.
\begin{equation}
    n_{\rm bubbles}=\mathcal{O}(1)\times e^{3N-B} < 1 \; .
\end{equation}
Both instanton solutions lead to bounds on the Hubble rate at the beginning of kination that are weaker than the constraints given by \eqref{eq:barrier_overcoming}. From \eqref{eq:fit_rho_br} we know that the energy-density of the Higgs at the end of the tachyonic phase is at least four orders of magnitude larger than the Hubble-scale energy density, $\mathcal{H}^4<10^{-4}\rho_{\rm tac}$, with ${\rho_{\rm tac}<V_{\rm eff}^{\rm 1 loop}(h_{\rm max})}$. Therefore, the classical stability constraint is much stronger than the constraint on the non-formation of bubbles from Hawking-Moss instantons, ${\mathcal{H}^4<\mathcal{O}(1)V_{\rm eff}^{\rm 1 loop}(h_{\rm max})}$. For Coleman-de Luccia solutions, the decay exponent is almost constant $B_{\rm CL}(\mathcal{H}) \simeq 1315$ and sufficiently large not to lead to the formation of bubbles in $\sim 10$ $e$-folds of heating. The lifetime of the metastable EW vacuum $\tau \simeq 10^{582} \text{ years}$ is comparable to the one computed in the SM alone, $\tau_{\rm SM} \simeq 10^{{983}^{+1410}_{-430}} \text{ years}$ \cite{Khoury:2021zao}.

%%%%%%%%%%%%%%%%%%%%%%%%%%%%%%%%%%%%%%%%%%%%%%%%%%%%%%%%%%%%%%%%%%%%%%%%%%%
\section{From Higgs instability to the hot Big Bang} \label{sec:reheating_constraints}
%%%%%%%%%%%%%%%%%%%%%%%%%%%%%%%%%%%%%%%%%%%%%%%%%%%%%%%%%%%%%%%%%%%%%%%%%%%

The explosive enhancement of the Higgs fluctuations in the tachyonically-unstable phase allows it to achieve a sufficient energy-density to heat the Universe in a few $e$-folds. Its energy density scales in a radiation-like fashion, thus growing over the kination background as $a^2$ and triggering a radiation-domination epoch following kination \cite{Bettoni:2018utf}. Additional constraints must be enforced if the Higgs field is identified with the reheaton field. Firstly, BBN sets a lower bound on the typical temperature of the thermalised plasma at the end of the heating phase: $T_{\rm ht} \gtrsim T_{\rm BBN}\approx 5 \text{ MeV}$ \cite{deSalas:2015glj, Hasegawa:2019jsa}. Whether the Higgs achieves a thermal spectrum by the end of the heating stage cannot be stated with confidence \cite{Micha:2004bv, Micha:2002ey, Micha:2003ws}. Nonetheless, a \textit{radiation temperature} can be always associated to the heated Universe \cite{Rubio:2017gty}
\begin{equation}\label{eq:temp_reheating}
T_{\rm ht} =\left(\frac{30\,\rho^{\rm ht}_{h}}{\pi^2 g_*^{\rm ht}}\right)^{1/4} \,,   
\end{equation}
with $g_*^{\rm ht}=106.75$ the SM number of relativistic degrees of freedom at energies above ${\cal O}(100) \text{ GeV}$ and $\rho^{\rm ht}_{h}=\rho^{\rm ht}_{\phi}$ the total energy-density of the Higgs at the end of the heating phase. The fitting formulas for $\rho^{\rm ht}_{h}(\lambda(\mu), \xi)$ in \cite{Laverda:2023uqv} allow us to estimate the heating temperature. We assume $\lambda(\mu)\approx\lambda_{\rm min}$ to be a good estimate of the Higgs self coupling close to the instability scale. The radiation time (i.e. the time at which the Higgs sector achieves a radiation-like equation of state) is set via the parametric formulas in \cite{Laverda:2023uqv} to typical values of a single-scalar-field heating scenario, i.e. $N_{\rm rad}\sim \mathcal{O}(1)$. In principle, the perturbative and non-perturbative interaction with additional SM degrees of freedom can favour the redistribution of energy among different species and modify the course of the heating stage \cite{Garcia-Bellido:2008ycs,Repond:2016sol}. However, any other production process of bosonic and fermionic fields is much less efficient than the Higgs tachyonic instability and the overall timeline of the heating process is defined exclusively by the Higgs energy density. In Appendix \ref{app:higgs_coupled_sm} we summarise in more detail the arguments why the presence of gauge bosons and heavy fermions does not modify noticeably neither the heating phase nor the vacuum stability constraints. 

\begin{figure}[tb]
\centering
\includegraphics[width=.65\textwidth]{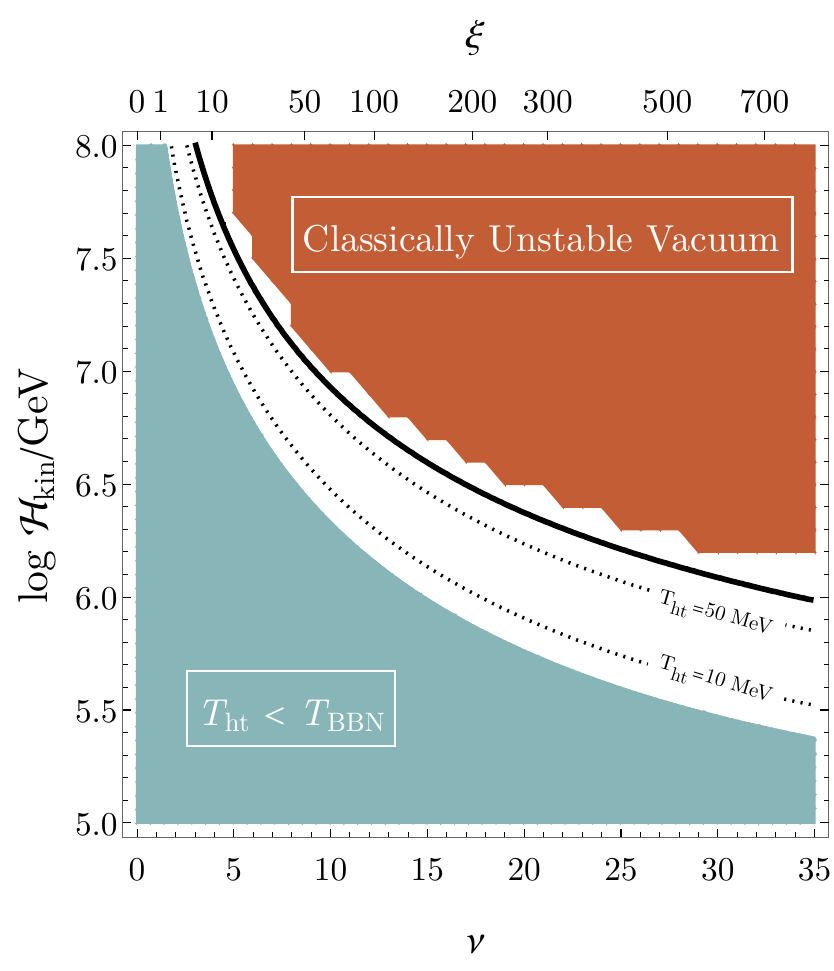}
\caption{Constraints on the model parameters $\nu=\sqrt{3\xi/2}$ and $\mathcal{H}_{\rm kin}$ given by the classical stability constraint for the one-loop running of $\lambda(\mu)$. The analytical criterion in  \eqref{eq:barrier_overcoming} is displayed as a black line. The red area is excluded due to the overcoming of the barrier in the numerical lattice simulations. The blue area is excluded by the bounds on the minimum reheating temperature. Dashed lines indicate some reference values of heating temperature $T_{\rm ht}$. We have set $m_t=171.3 \text{ GeV}$, which implies an instability scale of $\mu^{\rm 1loop}_{\Lambda}=3.5\times10^{9} \text{ GeV}$.} \label{fig:numerical_stability_1loop}
\end{figure}

As a prototypical case study, we perform a full scanning of the parameter space ${\nu \in [1, \; 35]}$, ${\mathcal{H}_{\rm kin} \in \left[ 10^{5} \text{ GeV}, \; 10^{8} \text{ GeV}\right]}$ with more than one thousand classical lattice simulations to identify the stable and unstable areas of the parameter space and compare them with the constraints on post-inflationary heating. We implement the non-minimally-coupled Higgs model in the lattice code \texttt{$\mathcal{C}osmo\mathcal{L}attice$} \cite{Figueroa:2020rrl, Figueroa:2021yhd} with the RGI effective potential in \eqref{eq:effective_potential} defined by the one-loop running of $\lambda(\mu)$ for $m_t=171.3 \text{ GeV}$. Because of the fast enhancement of the quantum modes amplitudes during the tachyonic phase, the classical 3+1-dimensional dynamics of fields on the lattice is a good description of the non-linearly interacting Higgs. The system is evolved for around two $e$-folds after the beginning of the kination phase, so to cover the full tachyonic instability phase. The details of the lattice setup are summarised in  Appendix \ref{app:lattice}.

In Figure~\ref{fig:numerical_stability_1loop}, the constraints on vacuum stability and heating are complementary in excluding most of the parameter space in the range under consideration. However, there exist a substantial difference between the two constraints: the stability of the Higgs vacuum lays at least one order of magnitude below the instability scale, while the heating constraint depends mostly on the scale of kination and excludes all values of $\mathcal{H}_{\rm kin}\lesssim10^{5.5} \text{ GeV}$. A larger viable portion of the parameter space can be achieved with instability scales greater than $\mu_{\Lambda}\gtrsim10^{9} \text{ GeV}$. Therefore, in the one-loop approximation, the Higgs can alone be responsible for the heating as long as the top quark mass is close to $171 \text{ GeV}$ or smaller, and the corresponding instability scale is higher than $10^9 \text{ GeV}$. 

Low inflationary scales $\mathcal{H}_{\rm kin}\lesssim 10^{5.5} \text{ GeV}$ are still compatible with a stable Higgs vacuum, but the Universe's post-inflationary heating mechanism has to be independent of the Higgs sector. If this is the case, the thermalisation of the Higgs with the heating sector can induce thermal corrections to the effective potential that could potentially rescue unstable patches of the Universe \cite{Bezrukov:2014ipa, Espinosa:2015qea}. This mechanism is obviously absent if the Higgs itself is tasked with heating the Universe.

A second constraint on the efficiency of the heating process is set by avoiding the overproduction of gravitational waves \cite{Bettoni:2021zhq}. Because of the typical high efficiency of the tachyonic production, this constraint is always satisfied in the parameter space shown in Figure \ref{fig:numerical_stability_1loop}.

%%%%%%%%%%%%%%%%%%%%%%%%%%%%%%%%%%%%%%%%%%%%%%%%%%%%%%%%%%%%%%%%%%%%%%%%%%%
\section{Improved constraints at three loops} \label{sec:stability_3loop}
%%%%%%%%%%%%%%%%%%%%%%%%%%%%%%%%%%%%%%%%%%%%%%%%%%%%%%%%%%%%%%%%%%%%%%%%%%%

\begin{figure}[tb]
\centering
\includegraphics[width=.65\textwidth]{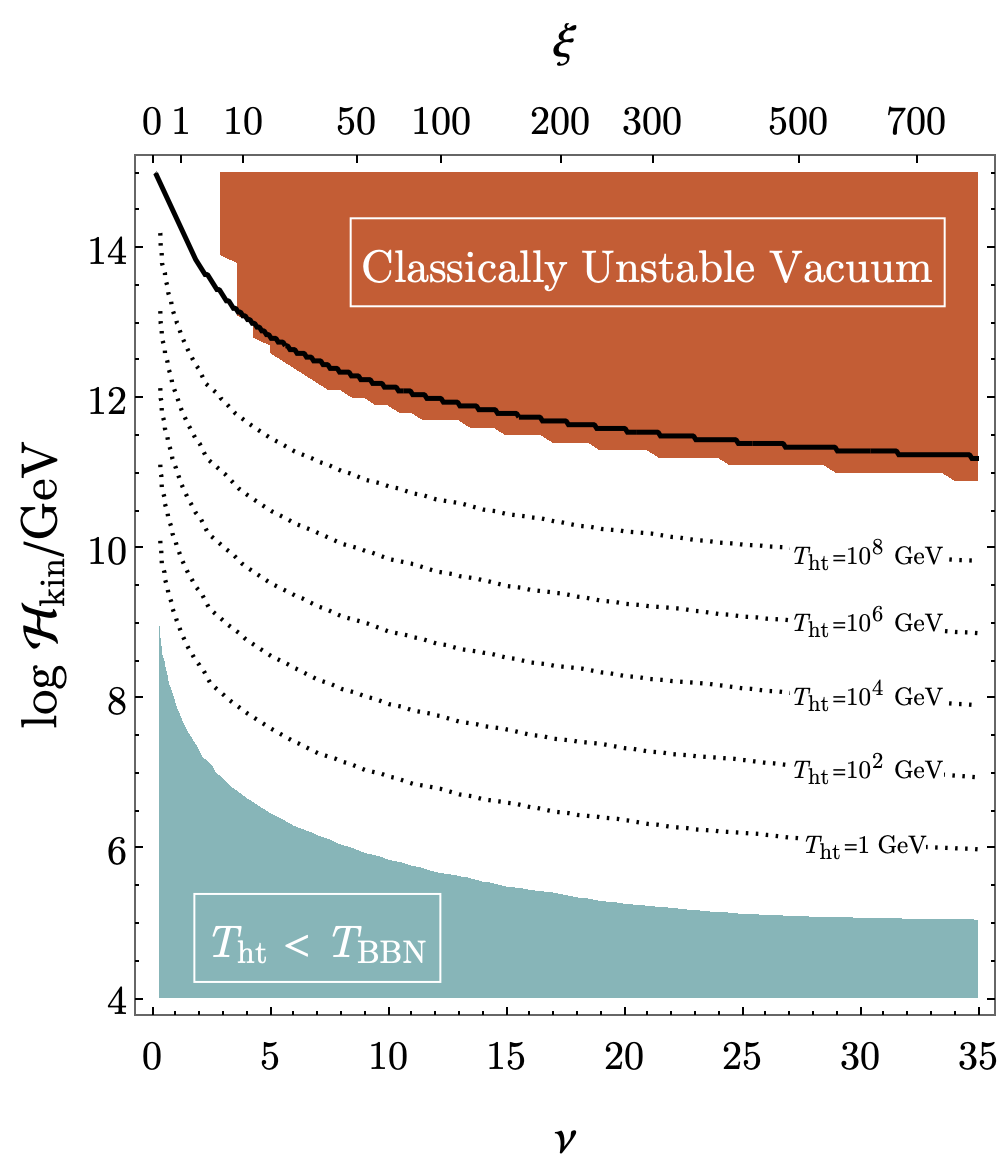}
\caption{Constraints on the parameters $\nu=\sqrt{3\xi/2}$ and $\mathcal{H}_{\rm kin}$ given by the classical stability constraint for the running of $\lambda(\mu)$ at three-loops. The red area corresponds to the unstable configurations as found via numerical simulations, with the black solid line separating the region of parameter space that fulfils the inequality \eqref{eq:barrier_overcoming} (lower part) from the one that does not (upper part). The blue area is excluded due to the constraints on the temperature at the end of the heating stage. Dashed lines indicate some reference values of heating temperature $T_{\rm ht}$. The top quark mass is set to $m_t=171.3 \text{ GeV}$, which implies an instability scale of $\mu^{\rm 3loop}_{\Lambda}=2.7\times10^{15} \text{ GeV}$.} \label{fig:numerical_stability_3loop}
\end{figure}

The simple one-loop analysis is capable of highlighting some of the essential features but shows two important drawbacks. Firstly, the instability scale is unavoidably low, thus limiting the parameter space that allows a successful achievement of heating before BBN. Secondly, the approximate running in \eqref{eq:lambda_running_approx} makes it difficult to recast the bounds we have found into constraints on the top quark mass at the EW scale, as our choice for the renormalisation scale $\mu$ depends on the non-minimal coupling parameter $\xi$ as well as on $\mathcal{H}_{\rm kin}$. Both points can be addressed by reformulating our analysis with the three-loop running of $\lambda$ and, more specifically, by approximating it around the instability scale as \cite{Bezrukov:2014bra}
\begin{equation} \label{eq:running_bezrukov}
    \lambda^{\rm 3loop}(\mu)=\lambda_0 + b \ln^2 \left[\frac{\mu}{q \, \rm M_P} \right]  \; ,
\end{equation}
where the parameters $\lambda_0$, $b$ and $q$ are given by
\begin{align}
    \lambda_0 &= 0.003297 ((m_h-126.13)-2 (m_t-171.5)) \; , \\
    q &= 0.3 \exp\left[(0.5(m_h - 126.13) - 0.03(m_t - 171.5))\right] \; , \\
    b &= 0.00002292 - 1.12524\times10^{-6}((m_h - 126.13) - 1.75912(m_t - 171.5)) \; ,
\end{align}
with a dependence on the Higgs and top quark masses $m_h$ and $m_t$ at the EW scale. This expression can be inserted directly into the effective potential \eqref{eq:effective_potential} in order to investigate the possible overtaking of the barrier. The equation $\rho_{\rm tac}=V_{\rm eff}^{\rm 3loop}(h_{\rm max})$ has to be solved numerically by first finding solutions to $\partial V_{\rm eff}^{\rm 3loop}/\partial h=0$, then evaluating $\lambda(h_{\rm min})$ and finally computing $\rho_{\rm tac}(\lambda(h_{\rm min}))$ and $V_{\rm eff}^{\rm 3loop}(h_{\rm max})$. The analysis of the parameter space is performed on a grid of points that covers the most interesting range of parameters. In analogy with \eqref{eq:barrier_overcoming}, the Higgs vacuum is stable if $\rho_{\rm tac}(\lambda(h_{\rm min}))<V_{\rm eff}^{\rm 3loop}(h_{\rm max})$. We adopt this criterion to separate with a black line in Figure~\ref{fig:numerical_stability_3loop} the absolutely unstable upper region of the parameter space from the stable region below the line.

\begin{figure}[tb]
\centering
\includegraphics[width=.8\textwidth]{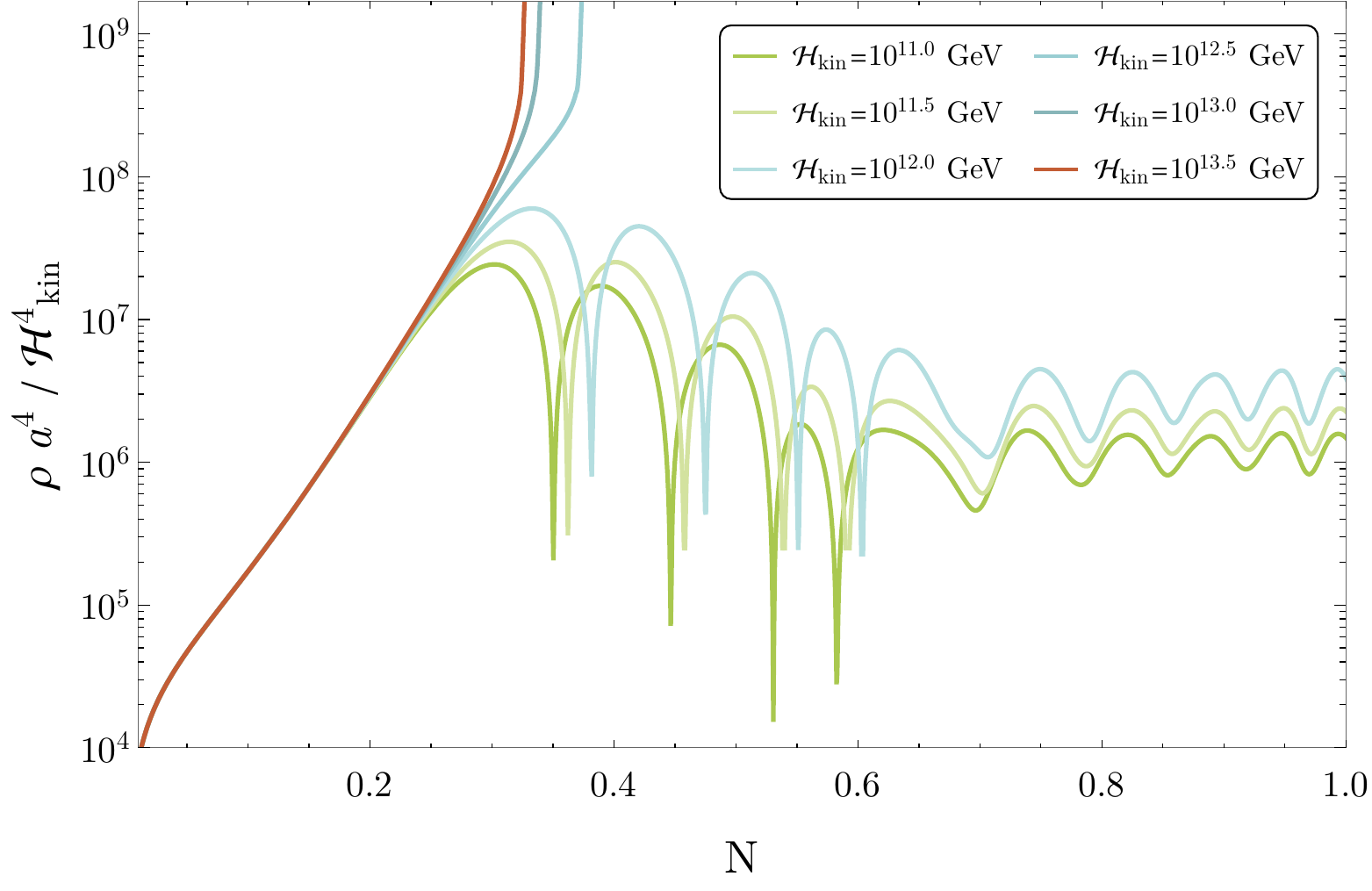}
\caption{Evolution of the lattice-averaged energy density of the Higgs field in the first $e$-fold of kination for different choices of inflationary scale $\mathcal{H}_{\rm kin}$ with a fixed value of $\nu=10$. The three-loop running of the Higgs self-coupling has been obtained setting $m_t=171.3 \text{ GeV}$. The total energy density is multiplied by a factor $a^4$ to highlight the radiation-like scaling. } \label{fig:energy_divergence}
\end{figure}

We scan numerically the parameter space $\nu \in [1, \, 35]$, $\mathcal{H}_{\rm kin} \in \left[ 10^{4} \text{ GeV}, \, 10^{15} \text{ GeV}\right]$ with a large number of short lattice simulations for a fiducial value of the top quark mass $m_t=171.3 \text{ GeV}$ and the three-loop running of $\lambda(\mu)$. The overall setup is the same as in Section \ref{sec:stability_1loop} and more details can be found in Appendix \ref{app:lattice}. Figure \ref{fig:energy_divergence} displays the different evolution of the lattice-averaged Higgs energy-densities in the stable and unstable scenario for six of our simulations. For $\mathcal{H}_{\rm kin}>10^{12} \text{ GeV}$, the tachyonically-amplified fluctuations overcome the potential barrier separating the two vacua and the field diverges rapidly to large field-values. In the stable case, the Higgs undergoes a non-linear evolution and eventually settles into an oscillatory dynamics in its almost-quartic potential. The results of the full scanning are contained in Figure~\ref{fig:numerical_stability_3loop}, where the classically unstable region is indicated in red and correctly identified by our estimates with the approximated three-loop running. Because of the statistical nature of the lattice approach which is inherited by the parametric formulas we are using, the analytic bound is to be understood as an indicative constraint. Random fluctuations in the simulations can lead to locally unstable patches of the Universe, even when the lattice-averaged energy-density would macroscopically indicate vacuum stability. The dashed lines in the same figure highlight the different radiation temperatures achieved at the end of the heating stage, as given by \eqref{eq:temp_reheating}. The comparison of our results with those in \cite{Opferkuch:2019zbd} indicates that there exist a much larger portion of parameter space that simultaneously allows for vacuum stability and successful heating, with heating temperatures much above $1 \text{ GeV}$ if the top quark mass is set to a value compatible with the latest measurements \cite{CMS:2023ebf, Myllymaki:2024uje}.

\begin{figure}[tb]
\centering
\includegraphics[width=.65\textwidth]{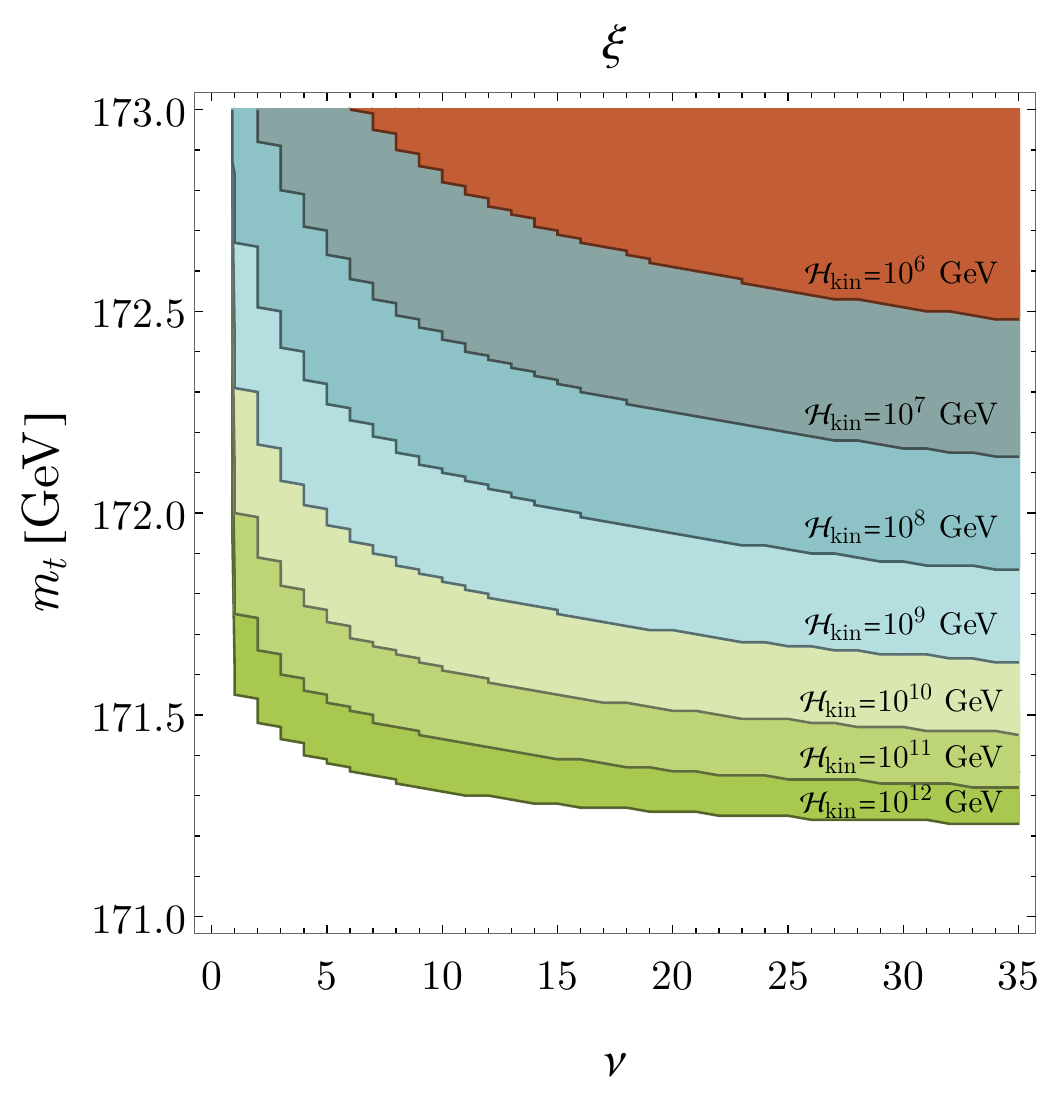}
\caption{Constraints on the mass of the top quark given by the classical stability as a function of $\nu=\sqrt{3\xi/2}$. The condition on the not-overcoming of the barrier in the effective potential is recast by keeping the top quark mass at the EW scale as a free parameter and finding numerical solutions to the equation $\rho_{\rm tac}<V_{\rm eff}^{\rm 3loop}(h_{\rm max})$. The coloured areas indicate the excluded unstable configurations, with different shades corresponding to different inflationary scales. Each region includes the coloured ones above it.}
\label{fig:constraints_top_mass}
\end{figure}

The comparison with numerical simulations indicates that the analytical estimates can describe sufficiently well the instability range to be used as constraining conditions on the mass of the top quark. Indeed, having at our disposal the running in \eqref{eq:running_bezrukov}, we can investigate the full space of model parameters $(m_t, \, \nu, \, \mathcal{H}_{\rm kin})$ and analyse which combinations of non-minimal coupling and top quark mass satisfy the stability criterion $\rho_{\rm tac}<V_{\rm eff}^{\rm 3loop}(h_{\rm max})$. Figure \ref{fig:constraints_top_mass} shows in different colours the areas that are excluded by vacuum instability. Since smaller top quark masses lead to higher instability scales, an inflationary phase terminating at low energies is overall favoured, while the more typical inflationary scales above $10^{10} \text{ GeV}$ \cite{Planck:2018jri} are allowed for masses somewhat below $171.5 \text{ GeV}$.  

\begin{figure}[tb]
\centering
\includegraphics[width=.65\textwidth]{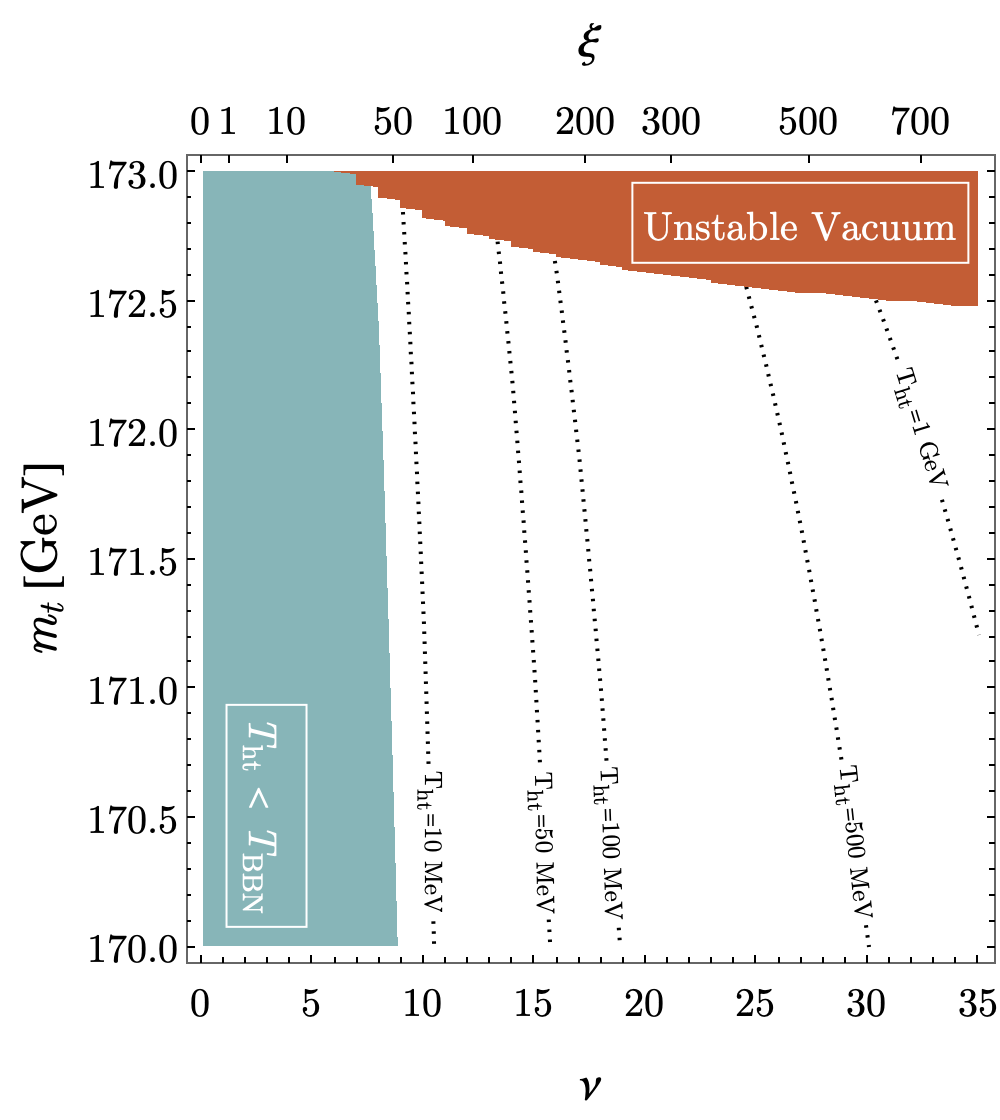}
\caption{Constraints on the parameters $\nu=\sqrt{3\xi/2}$ and $m_t$ given by the classical stability constraint for the running of $\lambda(\mu)$ at three-loops and an inflationary scale of ${\mathcal{H}_{\rm kin}=10^{6}} \text{ GeV}$. The red area corresponds to the unstable configurations as found via numerical solutions to ${\rho_{\rm tac}<V_{\rm eff}^{\rm 3loop}(h_{\rm max})}$. The blue area is excluded due to the constraints on the temperature at the end of the heating stage. Dotted lines give some indicative values of  $T_{\rm ht}$ \cite{deSalas:2015glj, Hasegawa:2019jsa}.} \label{fig:top_mass_constraints_3loop_1}
\end{figure}

\begin{figure}[tb]
\centering
\includegraphics[width=.65\textwidth]{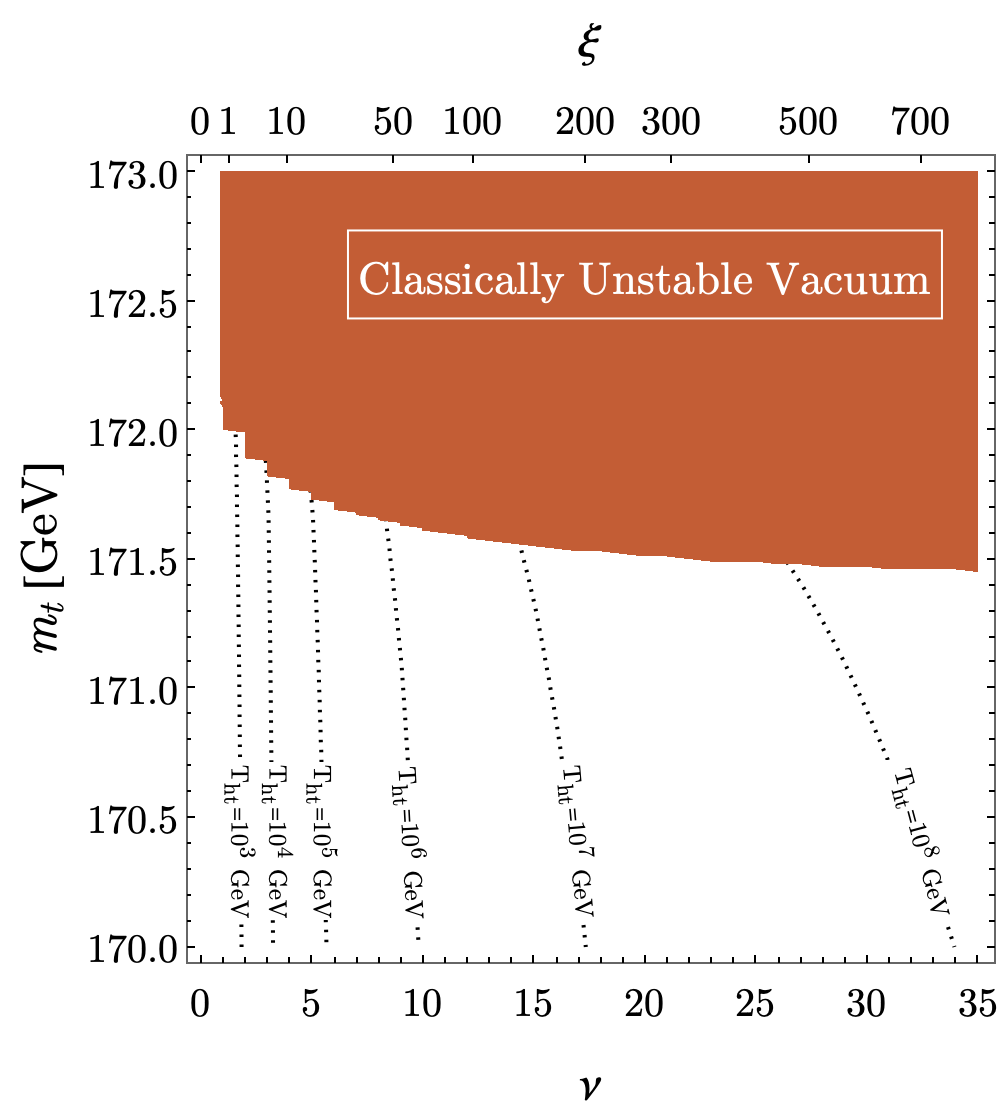}
\caption{Constraints on the parameters $\nu=\sqrt{3\xi/2}$ and $m_t$ given by the classical stability constraint for the running of $\lambda(\mu)$ at three-loops and an inflationary scale of $\mathcal{H}_{\rm kin}=10^{10}$ GeV. The red area corresponds to the unstable configurations as found via numerical solutions to the inequality ${\rho_{\rm tac}>V_{\rm eff}^{\rm 3loop}(h_{\rm max})}$. Dotted lines give some indicative values of the radiation temperature $T_{\rm ht}$. } \label{fig:top_mass_constraints_3loop_2}
\end{figure}

Figures \ref{fig:top_mass_constraints_3loop_1} and \ref{fig:top_mass_constraints_3loop_2} display how the criteria we have adopted can simultaneously constrain the space $(\nu, \, m_t)$ from the vacuum-stability and the heating-temperature point of view. The unstable region is shown in red, while BBN constraints are shown in blue. For a low inflationary scale $\mathcal{H}_{\rm kin}=10^{6}$ GeV (Figure \ref{fig:top_mass_constraints_3loop_1}), the heating process is not efficient, which tends to exclude low values of the non-minimal-coupling parameters and low masses of the top quark, i.e. higher instability scales. On the other hand, higher inflationary scales (Figure \ref{fig:top_mass_constraints_3loop_2}) lead to a more pronounced production of Higgs particles and to a faster (and hotter) end of the heating phase. However, in this scenario the instability constraint excludes a larger portion of the parameter space due to the more intense non-minimal interaction. We notice that the stability constraints do not affect the SM scenario without a non-minimal coupling coupling to curvature for $\nu \to 0$.

%%%%%%%%%%%%%%%%%%%%%%%%%%%%%%%%%%%%%%%%%%%%%%%%%%%%%%%%%%%%%%%%%%%%%%%%%%%
\section{Conclusions} \label{sec:conclusions}
%%%%%%%%%%%%%%%%%%%%%%%%%%%%%%%%%%%%%%%%%%%%%%%%%%%%%%%%%%%%%%%%%%%%%%%%%%%

The current measurements of the Higgs and the top quark masses at the Large Hadron Collider allow, within the experimental and theoretical uncertainties, for the instability of the electroweak vacuum at very high energies. In this work, we have presented an alternative history of the Higgs as a spectator field during a phase of kination following the end of inflation. Thanks to the double nature of the non-minimal coupling to curvature, the Higgs is safely stabilised during the inflationary epoch but undergoes a tachyonic instability during kination. The transition between the two phases acts effectively as a natural cosmic clock that triggers a copious non-perturbative production of Higgs particles and brings its typical amplitude close to the electroweak vacuum instability scale. With the help of the one- and three-loop running of the Higgs self-coupling, their approximations around the instability scale and a set of lattice-based parametric formulas specifying the maximum field displacement and Higgs energy density during kination, we were able to set significant constraints on the parameter space $(m_t, \, \nu, \, \mathcal{H}_{\rm kin})$. The SM vacuum remains stable whenever a barrier is formed and the tachyonically-enhanced Higgs energy density is smaller than the barrier's height. If we task the Higgs with heating the post-inflationary Universe, further regions of the parameter space can be excluded. These constraints are set by semi-analytical results and investigated in detail via numerical lattice simulations for a specific choice of top quark mass. Overall, our results allow for heating temperatures in the range $10^{-2}-10^9$ GeV if the top quark mass is set to a value $m_t=171.3 \text{ GeV}$ compatible with the latest measurements \cite{CMS:2023ebf, Myllymaki:2024uje}, significantly extending the parameter space in \cite{Opferkuch:2019zbd} and opening the gate to implementing potential EW baryogenesis mechanisms \cite{Shaposhnikov:1987tw,Wagner:2023vqw}. 
 
Our investigation, while adopting a relatively straightforward approach, demonstrates the potential for broader applications. Specifically, it can be easily extended to accommodate a generic stiff equation of state $w$, as that potentially appearing in several contexts and stages in the evolution of the very early Universe \cite{Gouttenoire:2021jhk}. Indeed, as long as the global equation of state evolves from $w \leq 1/3$ to $w > 1/3$, the development of a tachyonic spectator field becomes almost unavoidable, leading with it to a similar post-inflationary  dynamics. Moreover, our framework can be naturally extended to encompass more standard cosmological epochs, with the possible benefit of producing a stochastic background of primordial gravitational waves within the observable window of current and future ground and space-based experiments \cite{Punturo:2010zz,Harry:2006fi,LISA}. In particular, other types of non-minimal interactions with the gravitational background, such as a Gauss-Bonnet term, are expected to produce a similar phase transition between inflation and a radiation-domination epoch. 

As future work, it would be interesting to extend our understanding of vacuum stability during kination beyond the classical framework, performing, for instance, a comprehensive examination of quantum tunnelling effects through a statistical lens alongside a dedicated analysis of the thermalisation effects following the Higgs tachyonic amplification. 

\appendix

%%%%%%%%%%%%%%%%%%%%%%%%%%%%%%%%%%%%%%%%%%%%%%%%%%%%%%%%%%%%%%%%%%%%%%%%%%%
\section{Impact of gauge bosons and fermions} \label{app:higgs_coupled_sm}
%%%%%%%%%%%%%%%%%%%%%%%%%%%%%%%%%%%%%%%%%%%%%%%%%%%%%%%%%%%%%%%%%%%%%%%%%%%

In this Appendix, we summarise several results about the perturbative and non-perturbative production of SM fields during kination. The objective is to show that each decay channel is too inefficient to change substantially the history of heating after inflation or, in other words, that the Higgs alone is responsible for the achievement of a radiation-domination phase after a few $e$-folds of kination. Starting from the Higgs decay into fermions, the channel $h \to \Bar{t}t$ is kinematically blocked \cite{Enqvist:2013kaa} while the decay rate for $h \to \Bar{b}b$ is very small given the typical amplitude of the Higgs fluctuations and becomes comparable to the Hubble scale only after $\mathcal{O}(10)$ $e$-folds. Other perturbative decays into lighter fermions are even more suppressed due to the smallness of their Yukawa couplings. 

The perturbative decay into gauge bosons $h \to W W$ and $h \to Z Z$ is also blocked for energy scales above $10^2 \text{ GeV}$ \cite{Enqvist:2013kaa}. However, the Higgs can resonantly produce gauge bosons as it oscillates in its effective potential. This process is relatively inefficient when compared to the explosive tachyonic production of the Higgs \cite{Garcia-Bellido:1997hex, Garcia-Bellido:2001dqy}. Considering for simplicity a Higgs field oscillating uniformly in a quartic potential as $h(z)=\text{cn}(z, 1/\sqrt{2})$, the resonant production of bosonic particles happens through successive exponential amplifications. For $q=g^2/\lambda \gg 1$, as is typical for SM couplings, the Floquet index becomes asymptotically $\mu_k \to \mu_{\rm max} = 0.2377$ \cite{Greene:1997fu}. Even in the best-case scenario, the resonant amplification of the daughter field energy-density in each oscillation is proportional to $\sim \exp (2 \mu_{\rm max})$. It takes at least $\mathcal{O}(10)$ oscillations for the energy densities of the bosonic fields to become comparable to the Higgs energy density, as was seen with lattice simulations in \cite{Laverda:2023uqv}. Therefore, the non-perturbative production of gauge bosons does not interfere with the timeline of heating and with the constraints on vacuum stability. The same conclusion remains true if we consider the additional decay of gauge bosons into fermions \cite{Fan:2021otj, Kofman:1997yn}. In spite of these considerations, non-Abelian interactions could play, however, an important role in the thermalisation of the SM plasma, with boson scatterings and annihilations inducing a faster approach to equipartition \cite{Enqvist:2015sua,Bodeker:2007fw}.

Regarding the vacuum stability problem, a strong decay into fermions can slightly deplete the Higgs energy-density during the first semi-oscillation to the point of increasing its stability. This effect should be taken into account for a more precise estimate of the stability/instability regions of Figures \ref{fig:numerical_stability_1loop} and \ref{fig:numerical_stability_3loop}.

%%%%%%%%%%%%%%%%%%%%%%%%%%%%%%%%%%%%%%%%%%%%%%%%%%%%%%%%%%%%%%%%%%%%%%%%%%%
\section{The setup behind lattice simulations} \label{app:lattice}
%%%%%%%%%%%%%%%%%%%%%%%%%%%%%%%%%%%%%%%%%%%%%%%%%%%%%%%%%%%%%%%%%%%%%%%%%%%

The numerical analysis of the classical stability of the Higgs fluctuations has been performed using the  \texttt{$\mathcal{C}osmo\mathcal{L}attice$} code \cite{Figueroa:2021yhd, Figueroa:2016wxr}, a 3+1-dimensional lattice code capable of evolving interacting gauge fields in an expanding background. We implemented a single-scalar-field scenario via a Klein-Gordon equation that contains the curvature-dependent mass term in \eqref{eq:higgs_lagrangian}. The numerical RGI running of $\lambda(\mu)$ at one and three loop has been included in the effective potential for a specific choice of top quark mass $m_t=171.3 \text{ GeV}$.

The lattice parameters are set to ensure the stability and reliability of the output. In particular, we have set the number of lattice points per dimension $N=256$ so that all relevant modes are always well within the associated infrared (IR) and ultraviolet (UV) resolution in momentum space, since $\kappa_{\rm IR}=2\pi/L$, $\kappa_{\rm UV}= \sqrt{3} N \kappa_{\rm IR} / 2$, with $L=N \, \delta x$ and $\delta x = 4\pi \, \nu / N$ is the length of the side of a lattice cell. In particular, these quantities are set to properly cover the band of tachyonic momenta \cite{Laverda:2023uqv}. Following the results of the linear analysis of the Hubble-induced phase transition scenario of \cite{Bettoni:2019dcw}, we identify the smallest momentum in the tachyonic band with {$\kappa_{\text{IR}} = \mathcal{H}_{\rm kin}$, while the largest amplified momentum is set to be smaller than the lattice's UV momentum, i.e. $\sqrt{4\nu^2 - 1}\mathcal{H}_{\rm kin} \ll \kappa_{\text{UV}}$}, where we have set the scale factor to be $a_{\rm kin}=1$ at the beginning of kination. This condition implies a constraint on the minimum number of lattice sites $N>2\sqrt{4\nu^2 - 1}/\sqrt{3}$ which is always fulfilled in our simulations. 

The time-step variable is chosen according to the stability criterion $\delta t / \delta x \ll 1/\sqrt{d}$ \cite{Figueroa:2021yhd}, with $d=3$ the number of spatial dimensions and we set $\delta t=0.1$ for $\nu \geq 10$ and $\delta t=0.01$ for $\nu < 10$. The kination background expansion is obtained by fixing the equation-of-state parameter to $w=1$. The system is evolved via a symplectic 4th order Velocity-Verlet evolver since it guarantees stability and precision of the numerical solutions when the conservation of energy cannot be explicitly checked. The initial conditions for our lattice simulations are set as $h(0)=h'(0)=0$,
in agreement with the inflationary picture developed in Section \ref{sec:setup}, with fluctuations over this homogeneous background included as Gaussian random fields, as done customarily for systems with short classicalisation times \cite{Bettoni:2021zhq}. The resulting evolution is therefore deterministic up to a base seed that randomises the initial fluctuations. We choose to keep the base seed constant in all our simulations and make them exactly comparable. Since the system looses memory of the initial conditions soon after the development of the tachyonic instability, this choice does not influence the overall macroscopic evolution. However, a more robust but time-consuming approach would involve averaging the output of repeated simulations with random initial seed.

\acknowledgments
G.L. and J.R wish to thank Dario Bettoni, Andreas Mantziris and Matteo Piani for the interesting discussions during the preparation of this work. This work was partially supported by the Spanish Ministerio de Ciencia, Innovación y Universidades' project PID2022-139841NB-I00 (AEI/FEDER, UE). The numerical lattice simulations have been performed with the support of the Infraestrutura Nacional de Computa\c c\~ao Distribu\'ida (INCD) funded by the Funda\c c\~ao para a Ci\^encia e a Tecnologia (FCT) and FEDER under the project 01/SAICT/2016 nº 022153. G.L. (ORCID 0000-0002-4739-4946) is supported by a fellowship from ”la Caixa” Foundation (ID 100010434) with fellowship code LCF/BQ/DI21/11860024. G.~L. thanks also FCT for the financial support to the Center for Astrophysics and Gravitation-CENTRA, Instituto Superior T\'ecnico,  Universidade de Lisboa, through the Project No.~UIDB/00099/2020. J.R. (ORCID ID 0000-0001-7545-1533) is supported by a Ram\'on y Cajal contract of the Spanish Ministry of Science and Innovation with Ref.~RYC2020-028870-I.

\bibliographystyle{JHEP}
\bibliography{biblio.bib}

\providecommand{\href}[2]{#2}\begingroup\raggedright\begin{thebibliography}{100}

\bibitem{ATLAS:2012yve}
{\scshape ATLAS} collaboration, \emph{{Observation of a new particle in the search for the Standard Model Higgs boson with the ATLAS detector at the LHC}}, \href{https://doi.org/10.1016/j.physletb.2012.08.020}{\emph{Phys. Lett. B} {\bfseries 716} (2012) 1} [\href{https://arxiv.org/abs/1207.7214}{{\ttfamily 1207.7214}}].

\bibitem{CMS:2012qbp}
{\scshape CMS} collaboration, \emph{{Observation of a New Boson at a Mass of 125 GeV with the CMS Experiment at the LHC}}, \href{https://doi.org/10.1016/j.physletb.2012.08.021}{\emph{Phys. Lett. B} {\bfseries 716} (2012) 30} [\href{https://arxiv.org/abs/1207.7235}{{\ttfamily 1207.7235}}].

\bibitem{Tang:2013bz}
Y.~Tang, \emph{{Vacuum Stability in the Standard Model}}, \href{https://doi.org/10.1142/S0217732313300024}{\emph{Mod. Phys. Lett. A} {\bfseries 28} (2013) 1330002} [\href{https://arxiv.org/abs/1301.5812}{{\ttfamily 1301.5812}}].

\bibitem{Bezrukov:2014ina}
F.~Bezrukov and M.~Shaposhnikov, \emph{{Why should we care about the top quark Yukawa coupling?}}, \href{https://doi.org/10.1134/S1063776115030152}{\emph{J. Exp. Theor. Phys.} {\bfseries 120} (2015) 335} [\href{https://arxiv.org/abs/1411.1923}{{\ttfamily 1411.1923}}].

\bibitem{Hiller:2024zjp}
G.~Hiller, T.~H\"ohne, D.F.~Litim and T.~Steudtner, \emph{{Vacuum Stability in the Standard Model and Beyond}},  \href{https://arxiv.org/abs/2401.08811}{{\ttfamily 2401.08811}}.

\bibitem{Elias-Miro:2012eoi}
J.~Elias-Miro, J.R.~Espinosa, G.F.~Giudice, H.M.~Lee and A.~Strumia, \emph{{Stabilization of the Electroweak Vacuum by a Scalar Threshold Effect}}, \href{https://doi.org/10.1007/JHEP06(2012)031}{\emph{JHEP} {\bfseries 06} (2012) 031} [\href{https://arxiv.org/abs/1203.0237}{{\ttfamily 1203.0237}}].

\bibitem{Espinosa:2013lma}
J.R.~Espinosa, \emph{{Vacuum Stability and the Higgs Boson}}, \href{https://doi.org/10.22323/1.187.0010}{\emph{PoS} {\bfseries LATTICE2013} (2014) 010} [\href{https://arxiv.org/abs/1311.1970}{{\ttfamily 1311.1970}}].

\bibitem{Branchina:2013jra}
V.~Branchina and E.~Messina, \emph{{Stability, Higgs Boson Mass and New Physics}}, \href{https://doi.org/10.1103/PhysRevLett.111.241801}{\emph{Phys. Rev. Lett.} {\bfseries 111} (2013) 241801} [\href{https://arxiv.org/abs/1307.5193}{{\ttfamily 1307.5193}}].

\bibitem{Branchina:2014rva}
V.~Branchina, E.~Messina and M.~Sher, \emph{{Lifetime of the electroweak vacuum and sensitivity to Planck scale physics}}, \href{https://doi.org/10.1103/PhysRevD.91.013003}{\emph{Phys. Rev. D} {\bfseries 91} (2015) 013003} [\href{https://arxiv.org/abs/1408.5302}{{\ttfamily 1408.5302}}].

\bibitem{Branchina:2014usa}
V.~Branchina, E.~Messina and A.~Platania, \emph{{Top mass determination, Higgs inflation, and vacuum stability}}, \href{https://doi.org/10.1007/JHEP09(2014)182}{\emph{JHEP} {\bfseries 09} (2014) 182} [\href{https://arxiv.org/abs/1407.4112}{{\ttfamily 1407.4112}}].

\bibitem{Domenech:2020yjf}
G.~Dom\`enech, M.~Goodsell and C.~Wetterich, \emph{{Neutrino masses, vacuum stability and quantum gravity prediction for the mass of the top quark}}, \href{https://doi.org/10.1007/JHEP01(2021)180}{\emph{JHEP} {\bfseries 01} (2021) 180} [\href{https://arxiv.org/abs/2008.04310}{{\ttfamily 2008.04310}}].

\bibitem{Bezrukov:2014ipa}
F.~Bezrukov, J.~Rubio and M.~Shaposhnikov, \emph{{Living beyond the edge: Higgs inflation and vacuum metastability}}, \href{https://doi.org/10.1103/PhysRevD.92.083512}{\emph{Phys. Rev. D} {\bfseries 92} (2015) 083512} [\href{https://arxiv.org/abs/1412.3811}{{\ttfamily 1412.3811}}].

\bibitem{Markkanen:2018pdo}
T.~Markkanen, A.~Rajantie and S.~Stopyra, \emph{{Cosmological Aspects of Higgs Vacuum Metastability}}, \href{https://doi.org/10.3389/fspas.2018.00040}{\emph{Front. Astron. Space Sci.} {\bfseries 5} (2018) 40} [\href{https://arxiv.org/abs/1809.06923}{{\ttfamily 1809.06923}}].

\bibitem{Markkanen:2018bfx}
T.~Markkanen, S.~Nurmi, A.~Rajantie and S.~Stopyra, \emph{{The 1-loop effective potential for the Standard Model in curved spacetime}}, \href{https://doi.org/10.1007/JHEP06(2018)040}{\emph{JHEP} {\bfseries 06} (2018) 040} [\href{https://arxiv.org/abs/1804.02020}{{\ttfamily 1804.02020}}].

\bibitem{ATLAS:2018fwq}
{\scshape ATLAS} collaboration, \emph{{Measurement of the top quark mass in the $t\bar{t}\rightarrow $ lepton+jets channel from $\sqrt{s}=8$ TeV ATLAS data and combination with previous results}}, \href{https://doi.org/10.1140/epjc/s10052-019-6757-9}{\emph{Eur. Phys. J. C} {\bfseries 79} (2019) 290} [\href{https://arxiv.org/abs/1810.01772}{{\ttfamily 1810.01772}}].

\bibitem{CMS:2015lbj}
{\scshape CMS} collaboration, \emph{{Measurement of the top quark mass using proton-proton data at ${\sqrt{(s)}}$ = 7 and 8 TeV}}, \href{https://doi.org/10.1103/PhysRevD.93.072004}{\emph{Phys. Rev. D} {\bfseries 93} (2016) 072004} [\href{https://arxiv.org/abs/1509.04044}{{\ttfamily 1509.04044}}].

\bibitem{CDF:2016vzt}
{\scshape CDF, D0} collaboration, \emph{{Combination of CDF and D0 results on the mass of the top quark using up $9.7\:{\rm fb}^{-1}$ at the Tevatron}},  \href{https://arxiv.org/abs/1608.01881}{{\ttfamily 1608.01881}}.

\bibitem{CMS:2018quc}
{\scshape CMS} collaboration, \emph{{Measurement of the top quark mass with lepton+jets final states using $\mathrm {p}$ $\mathrm {p}$ collisions at $\sqrt{s}=13\,\text {TeV} $}}, \href{https://doi.org/10.1140/epjc/s10052-018-6332-9}{\emph{Eur. Phys. J. C} {\bfseries 78} (2018) 891} [\href{https://arxiv.org/abs/1805.01428}{{\ttfamily 1805.01428}}].

\bibitem{CMS:2023ebf}
{\scshape CMS} collaboration, \emph{{Measurement of the top quark mass using a profile likelihood approach with the lepton~+~jets final states in proton\textendash{}proton collisions at $\sqrt{s}=13\,\text {Te}\hspace{-.08em}\text {V} $}}, \href{https://doi.org/10.1140/epjc/s10052-023-12050-4}{\emph{Eur. Phys. J. C} {\bfseries 83} (2023) 963} [\href{https://arxiv.org/abs/2302.01967}{{\ttfamily 2302.01967}}].

\bibitem{Bierlich:2022pfr}
C.~Bierlich et~al., \emph{{A comprehensive guide to the physics and usage of PYTHIA 8.3}}, \href{https://doi.org/10.21468/SciPostPhysCodeb.8}{\emph{SciPost Phys. Codeb.} {\bfseries 2022} (2022) 8} [\href{https://arxiv.org/abs/2203.11601}{{\ttfamily 2203.11601}}].

\bibitem{Bellm:2015jjp}
J.~Bellm et~al., \emph{{Herwig 7.0/Herwig++ 3.0 release note}}, \href{https://doi.org/10.1140/epjc/s10052-016-4018-8}{\emph{Eur. Phys. J. C} {\bfseries 76} (2016) 196} [\href{https://arxiv.org/abs/1512.01178}{{\ttfamily 1512.01178}}].

\bibitem{Myllymaki:2024uje}
{\scshape ATLAS, CMS} collaboration, \emph{{Top mass measurements}},  in \emph{{16th International Workshop on Top Quark Physics}}, 1, 2024 [\href{https://arxiv.org/abs/2401.04824}{{\ttfamily 2401.04824}}].

\bibitem{CMS:2019esx}
{\scshape CMS} collaboration, \emph{{Measurement of $\mathrm{t\bar t}$ normalised multi-differential cross sections in pp collisions at $\sqrt s=13$ TeV, and simultaneous determination of the strong coupling strength, top quark pole mass, and parton distribution functions}}, \href{https://doi.org/10.1140/epjc/s10052-020-7917-7}{\emph{Eur. Phys. J. C} {\bfseries 80} (2020) 658} [\href{https://arxiv.org/abs/1904.05237}{{\ttfamily 1904.05237}}].

\bibitem{Steingasser:2023ugv}
T.~Steingasser and D.I.~Kaiser, \emph{{Higgs potential criticality beyond the Standard Model}}, \href{https://doi.org/10.1103/PhysRevD.108.095035}{\emph{Phys. Rev. D} {\bfseries 108} (2023) 095035} [\href{https://arxiv.org/abs/2307.10361}{{\ttfamily 2307.10361}}].

\bibitem{Espinosa:2015qea}
J.R.~Espinosa, G.F.~Giudice, E.~Morgante, A.~Riotto, L.~Senatore, A.~Strumia et~al., \emph{{The cosmological Higgstory of the vacuum instability}}, \href{https://doi.org/10.1007/JHEP09(2015)174}{\emph{JHEP} {\bfseries 09} (2015) 174} [\href{https://arxiv.org/abs/1505.04825}{{\ttfamily 1505.04825}}].

\bibitem{Strumia:2022kez}
A.~Strumia and N.~Tetradis, \emph{{Higgstory repeats itself}}, \href{https://doi.org/10.1007/JHEP09(2022)203}{\emph{JHEP} {\bfseries 09} (2022) 203} [\href{https://arxiv.org/abs/2207.00299}{{\ttfamily 2207.00299}}].

\bibitem{Figueroa:2016dsc}
D.G.~Figueroa and C.T.~Byrnes, \emph{{The Standard Model Higgs as the origin of the hot Big Bang}}, \href{https://doi.org/10.1016/j.physletb.2017.01.059}{\emph{Phys. Lett. B} {\bfseries 767} (2017) 272} [\href{https://arxiv.org/abs/1604.03905}{{\ttfamily 1604.03905}}].

\bibitem{Nakama:2018gll}
T.~Nakama and J.~Yokoyama, \emph{{Reheating through the Higgs amplified by spinodal instabilities and gravitational creation of gravitons}}, \href{https://doi.org/10.1093/ptep/ptz014}{\emph{PTEP} {\bfseries 2019} (2019) 033E02} [\href{https://arxiv.org/abs/1803.07111}{{\ttfamily 1803.07111}}].

\bibitem{Dimopoulos:2018wfg}
K.~Dimopoulos and T.~Markkanen, \emph{{Non-minimal gravitational reheating during kination}}, \href{https://doi.org/10.1088/1475-7516/2018/06/021}{\emph{JCAP} {\bfseries 06} (2018) 021} [\href{https://arxiv.org/abs/1803.07399}{{\ttfamily 1803.07399}}].

\bibitem{Bettoni:2018utf}
D.~Bettoni and J.~Rubio, \emph{{Quintessential Affleck-Dine baryogenesis with non-minimal couplings}}, \href{https://doi.org/10.1016/j.physletb.2018.07.046}{\emph{Phys. Lett. B} {\bfseries 784} (2018) 122} [\href{https://arxiv.org/abs/1805.02669}{{\ttfamily 1805.02669}}].

\bibitem{Bettoni:2018pbl}
D.~Bettoni, G.~Dom\`enech and J.~Rubio, \emph{{Gravitational waves from global cosmic strings in quintessential inflation}}, \href{https://doi.org/10.1088/1475-7516/2019/02/034}{\emph{JCAP} {\bfseries 02} (2019) 034} [\href{https://arxiv.org/abs/1810.11117}{{\ttfamily 1810.11117}}].

\bibitem{Bettoni:2019dcw}
D.~Bettoni and J.~Rubio, \emph{{Hubble-induced phase transitions: Walls are not forever}}, \href{https://doi.org/10.1088/1475-7516/2020/01/002}{\emph{JCAP} {\bfseries 01} (2020) 002} [\href{https://arxiv.org/abs/1911.03484}{{\ttfamily 1911.03484}}].

\bibitem{Bettoni:2021zhq}
D.~Bettoni, A.~Lopez-Eiguren and J.~Rubio, \emph{{Hubble-induced phase transitions on the lattice with applications to Ricci reheating}}, \href{https://doi.org/10.1088/1475-7516/2022/01/002}{\emph{JCAP} {\bfseries 01} (2022) 002} [\href{https://arxiv.org/abs/2107.09671}{{\ttfamily 2107.09671}}].

\bibitem{Laverda:2023uqv}
G.~Laverda and J.~Rubio, \emph{{Ricci reheating reloaded}}, \href{https://doi.org/10.1088/1475-7516/2024/03/033}{\emph{JCAP} {\bfseries 03} (2024) 033} [\href{https://arxiv.org/abs/2307.03774}{{\ttfamily 2307.03774}}].

\bibitem{Bettoni:2021qfs}
D.~Bettoni and J.~Rubio, \emph{{Quintessential Inflation: A Tale of Emergent and Broken Symmetries}}, \href{https://doi.org/10.3390/galaxies10010022}{\emph{Galaxies} {\bfseries 10} (2022) 22} [\href{https://arxiv.org/abs/2112.11948}{{\ttfamily 2112.11948}}].

\bibitem{Figueroa:2015rqa}
D.G.~Figueroa, J.~Garcia-Bellido and F.~Torrenti, \emph{{Decay of the standard model Higgs field after inflation}}, \href{https://doi.org/10.1103/PhysRevD.92.083511}{\emph{Phys. Rev. D} {\bfseries 92} (2015) 083511} [\href{https://arxiv.org/abs/1504.04600}{{\ttfamily 1504.04600}}].

\bibitem{Figueroa:2017slm}
D.G.~Figueroa, A.~Rajantie and F.~Torrenti, \emph{{Higgs field-curvature coupling and postinflationary vacuum instability}}, \href{https://doi.org/10.1103/PhysRevD.98.023532}{\emph{Phys. Rev. D} {\bfseries 98} (2018) 023532} [\href{https://arxiv.org/abs/1709.00398}{{\ttfamily 1709.00398}}].

\bibitem{Figueroa:2016wxr}
D.G.~Figueroa and F.~Torrenti, \emph{{Parametric resonance in the early Universe\textemdash{}a fitting analysis}}, \href{https://doi.org/10.1088/1475-7516/2017/02/001}{\emph{JCAP} {\bfseries 02} (2017) 001} [\href{https://arxiv.org/abs/1609.05197}{{\ttfamily 1609.05197}}].

\bibitem{Herranen:2015ima}
M.~Herranen, T.~Markkanen, S.~Nurmi and A.~Rajantie, \emph{{Spacetime curvature and Higgs stability after inflation}}, \href{https://doi.org/10.1103/PhysRevLett.115.241301}{\emph{Phys. Rev. Lett.} {\bfseries 115} (2015) 241301} [\href{https://arxiv.org/abs/1506.04065}{{\ttfamily 1506.04065}}].

\bibitem{Mantziris:2020rzh}
A.~Mantziris, T.~Markkanen and A.~Rajantie, \emph{{Vacuum decay constraints on the Higgs curvature coupling from inflation}}, \href{https://doi.org/10.1088/1475-7516/2021/03/077}{\emph{JCAP} {\bfseries 03} (2021) 077} [\href{https://arxiv.org/abs/2011.03763}{{\ttfamily 2011.03763}}].

\bibitem{Mantziris:2021oah}
A.~Mantziris, \emph{{Cosmological implications of EW vacuum instability: constraints on the Higgs-curvature coupling from inflation}}, \href{https://doi.org/10.22323/1.398.0127}{\emph{PoS} {\bfseries EPS-HEP2021} (2022) 127} [\href{https://arxiv.org/abs/2111.02464}{{\ttfamily 2111.02464}}].

\bibitem{Mantziris:2021zox}
A.~Mantziris, \emph{{On the cosmological implications of the electroweak vacuum instability: constraining the non-minimal coupling with inflation}}, \href{https://doi.org/10.1088/1742-6596/2156/1/012239}{\emph{J. Phys. Conf. Ser.} {\bfseries 2156} (2021) 012239} [\href{https://arxiv.org/abs/2111.02497}{{\ttfamily 2111.02497}}].

\bibitem{Mantziris:2022bfe}
A.~Mantziris, \emph{{Ending inflation with a bang: Higgs vacuum decay in $R^2$ gravity}}, \href{https://doi.org/10.22323/1.414.0114}{\emph{PoS} {\bfseries ICHEP2022} (2022) 114} [\href{https://arxiv.org/abs/2211.09244}{{\ttfamily 2211.09244}}].

\bibitem{Mantziris:2022fuu}
A.~Mantziris, T.~Markkanen and A.~Rajantie, \emph{{The effective Higgs potential and vacuum decay in Starobinsky inflation}}, \href{https://doi.org/10.1088/1475-7516/2022/10/073}{\emph{JCAP} {\bfseries 10} (2022) 073} [\href{https://arxiv.org/abs/2207.00696}{{\ttfamily 2207.00696}}].

\bibitem{Mantziris:2023xsp}
A.~Mantziris, \emph{{Higgs vacuum metastability in $R+R^2$ gravity}},  in \emph{{40th Conference on Recent Developments in High Energy Physics and Cosmology}}, 8, 2023 [\href{https://arxiv.org/abs/2308.00779}{{\ttfamily 2308.00779}}].

\bibitem{Kohri:2016wof}
K.~Kohri and H.~Matsui, \emph{{Higgs vacuum metastability in primordial inflation, preheating, and reheating}}, \href{https://doi.org/10.1103/PhysRevD.94.103509}{\emph{Phys. Rev. D} {\bfseries 94} (2016) 103509} [\href{https://arxiv.org/abs/1602.02100}{{\ttfamily 1602.02100}}].

\bibitem{Postma:2017hbk}
M.~Postma and J.~van~de Vis, \emph{{Electroweak stability and non-minimal coupling}}, \href{https://doi.org/10.1088/1475-7516/2017/05/004}{\emph{JCAP} {\bfseries 05} (2017) 004} [\href{https://arxiv.org/abs/1702.07636}{{\ttfamily 1702.07636}}].

\bibitem{Ema:2017loe}
Y.~Ema, M.~Karciauskas, O.~Lebedev and M.~Zatta, \emph{{Early Universe Higgs dynamics in the presence of the Higgs-inflaton and non-minimal Higgs-gravity couplings}}, \href{https://doi.org/10.1088/1475-7516/2017/06/054}{\emph{JCAP} {\bfseries 06} (2017) 054} [\href{https://arxiv.org/abs/1703.04681}{{\ttfamily 1703.04681}}].

\bibitem{Enqvist:2016mqj}
K.~Enqvist, M.~Karciauskas, O.~Lebedev, S.~Rusak and M.~Zatta, \emph{{Postinflationary vacuum instability and Higgs-inflaton couplings}}, \href{https://doi.org/10.1088/1475-7516/2016/11/025}{\emph{JCAP} {\bfseries 11} (2016) 025} [\href{https://arxiv.org/abs/1608.08848}{{\ttfamily 1608.08848}}].

\bibitem{Ema:2016kpf}
Y.~Ema, K.~Mukaida and K.~Nakayama, \emph{{Fate of Electroweak Vacuum during Preheating}}, \href{https://doi.org/10.1088/1475-7516/2016/10/043}{\emph{JCAP} {\bfseries 10} (2016) 043} [\href{https://arxiv.org/abs/1602.00483}{{\ttfamily 1602.00483}}].

\bibitem{Bezrukov:2014bra}
F.~Bezrukov and M.~Shaposhnikov, \emph{{Higgs inflation at the critical point}}, \href{https://doi.org/10.1016/j.physletb.2014.05.074}{\emph{Phys. Lett. B} {\bfseries 734} (2014) 249} [\href{https://arxiv.org/abs/1403.6078}{{\ttfamily 1403.6078}}].

\bibitem{Figueroa:2020rrl}
D.G.~Figueroa, A.~Florio, F.~Torrenti and W.~Valkenburg, \emph{{The art of simulating the early Universe -- Part I}}, \href{https://doi.org/10.1088/1475-7516/2021/04/035}{\emph{JCAP} {\bfseries 04} (2021) 035} [\href{https://arxiv.org/abs/2006.15122}{{\ttfamily 2006.15122}}].

\bibitem{Figueroa:2021yhd}
D.G.~Figueroa, A.~Florio, F.~Torrenti and W.~Valkenburg, \emph{{CosmoLattice: A modern code for lattice simulations of scalar and gauge field dynamics in an expanding universe}}, \href{https://doi.org/10.1016/j.cpc.2022.108586}{\emph{Comput. Phys. Commun.} {\bfseries 283} (2023) 108586} [\href{https://arxiv.org/abs/2102.01031}{{\ttfamily 2102.01031}}].

\bibitem{Birrell:1982ix}
N.D.~Birrell and P.C.W.~Davies, \emph{{Quantum Fields in Curved Space}}, Cambridge Monographs on Mathematical Physics, Cambridge Univ. Press, Cambridge, UK (2, 1984), \href{https://doi.org/10.1017/CBO9780511622632}{10.1017/CBO9780511622632}.

\bibitem{Mukhanov:2007zz}
V.~Mukhanov and S.~Winitzki, \emph{{Introduction to quantum effects in gravity}}, Cambridge University Press (6, 2007).

\bibitem{Apers:2022cyl}
F.~Apers, J.P.~Conlon, M.~Mosny and F.~Revello, \emph{{Kination, meet Kasner: on the asymptotic cosmology of string compactifications}}, \href{https://doi.org/10.1007/JHEP08(2023)156}{\emph{JHEP} {\bfseries 08} (2023) 156} [\href{https://arxiv.org/abs/2212.10293}{{\ttfamily 2212.10293}}].

\bibitem{Apers:2024ffe}
F.~Apers, J.P.~Conlon, E.J.~Copeland, M.~Mosny and F.~Revello, \emph{{String Theory and the First Half of the Universe}},  \href{https://arxiv.org/abs/2401.04064}{{\ttfamily 2401.04064}}.

\bibitem{Revello:2023hro}
F.~Revello, \emph{{Attractive (s)axions: cosmological trackers at the boundary of moduli space}},  \href{https://arxiv.org/abs/2311.12429}{{\ttfamily 2311.12429}}.

\bibitem{Conlon:2022pnx}
J.P.~Conlon and F.~Revello, \emph{{Catch-me-if-you-can: the overshoot problem and the weak/inflation hierarchy}}, \href{https://doi.org/10.1007/JHEP11(2022)155}{\emph{JHEP} {\bfseries 11} (2022) 155} [\href{https://arxiv.org/abs/2207.00567}{{\ttfamily 2207.00567}}].

\bibitem{Wetterich:1987fm}
C.~Wetterich, \emph{{Cosmology and the Fate of Dilatation Symmetry}}, \href{https://doi.org/10.1016/0550-3213(88)90193-9}{\emph{Nucl. Phys. B} {\bfseries 302} (1988) 668} [\href{https://arxiv.org/abs/1711.03844}{{\ttfamily 1711.03844}}].

\bibitem{Wetterich:1994bg}
C.~Wetterich, \emph{{The Cosmon model for an asymptotically vanishing time dependent cosmological 'constant'}}, {\emph{Astron. Astrophys.} {\bfseries 301} (1995) 321} [\href{https://arxiv.org/abs/hep-th/9408025}{{\ttfamily hep-th/9408025}}].

\bibitem{Peebles:1998qn}
P.J.E.~Peebles and A.~Vilenkin, \emph{{Quintessential inflation}}, \href{https://doi.org/10.1103/PhysRevD.59.063505}{\emph{Phys. Rev.} {\bfseries D59} (1999) 063505} [\href{https://arxiv.org/abs/astro-ph/9810509}{{\ttfamily astro-ph/9810509}}].

\bibitem{Spokoiny:1993kt}
B.~Spokoiny, \emph{{Deflationary universe scenario}}, \href{https://doi.org/10.1016/0370-2693(93)90155-B}{\emph{Phys. Lett.} {\bfseries B315} (1993) 40} [\href{https://arxiv.org/abs/gr-qc/9306008}{{\ttfamily gr-qc/9306008}}].

\bibitem{Brax:2005uf}
P.~Brax and J.~Martin, \emph{{Coupling quintessence to inflation in supergravity}}, \href{https://doi.org/10.1103/PhysRevD.71.063530}{\emph{Phys. Rev.} {\bfseries D71} (2005) 063530} [\href{https://arxiv.org/abs/astro-ph/0502069}{{\ttfamily astro-ph/0502069}}].

\bibitem{BuenoSanchez:2007jxm}
J.C.~Bueno~Sanchez and K.~Dimopoulos, \emph{{Curvaton reheating allows TeV Hubble scale in NO inflation}}, \href{https://doi.org/10.1088/1475-7516/2007/11/007}{\emph{JCAP} {\bfseries 11} (2007) 007} [\href{https://arxiv.org/abs/0707.3967}{{\ttfamily 0707.3967}}].

\bibitem{Wetterich:2013jsa}
C.~Wetterich, \emph{{Variable gravity Universe}}, \href{https://doi.org/10.1103/PhysRevD.89.024005}{\emph{Phys. Rev. D} {\bfseries 89} (2014) 024005} [\href{https://arxiv.org/abs/1308.1019}{{\ttfamily 1308.1019}}].

\bibitem{Wetterich:2014gaa}
C.~Wetterich, \emph{{Inflation, quintessence, and the origin of mass}}, \href{https://doi.org/10.1016/j.nuclphysb.2015.05.019}{\emph{Nucl. Phys. B} {\bfseries 897} (2015) 111} [\href{https://arxiv.org/abs/1408.0156}{{\ttfamily 1408.0156}}].

\bibitem{Hossain:2014xha}
M.W.~Hossain, R.~Myrzakulov, M.~Sami and E.N.~Saridakis, \emph{{Variable gravity: A suitable framework for quintessential inflation}}, \href{https://doi.org/10.1103/PhysRevD.90.023512}{\emph{Phys. Rev.} {\bfseries D90} (2014) 023512} [\href{https://arxiv.org/abs/1402.6661}{{\ttfamily 1402.6661}}].

\bibitem{Agarwal:2017wxo}
A.~Agarwal, R.~Myrzakulov, M.~Sami and N.K.~Singh, \emph{{Quintessential inflation in a thawing realization}}, \href{https://doi.org/10.1016/j.physletb.2017.04.066}{\emph{Phys. Lett.} {\bfseries B770} (2017) 200} [\href{https://arxiv.org/abs/1708.00156}{{\ttfamily 1708.00156}}].

\bibitem{Geng:2017mic}
C.-Q.~Geng, C.-C.~Lee, M.~Sami, E.N.~Saridakis and A.A.~Starobinsky, \emph{{Observational constraints on successful model of quintessential Inflation}}, \href{https://doi.org/10.1088/1475-7516/2017/06/011}{\emph{JCAP} {\bfseries 1706} (2017) 011} [\href{https://arxiv.org/abs/1705.01329}{{\ttfamily 1705.01329}}].

\bibitem{Dimopoulos:2017zvq}
K.~Dimopoulos and C.~Owen, \emph{{Quintessential Inflation with $\alpha$-attractors}}, \href{https://doi.org/10.1088/1475-7516/2017/06/027}{\emph{JCAP} {\bfseries 06} (2017) 027} [\href{https://arxiv.org/abs/1703.00305}{{\ttfamily 1703.00305}}].

\bibitem{Rubio:2017gty}
J.~Rubio and C.~Wetterich, \emph{{Emergent scale symmetry: Connecting inflation and dark energy}}, \href{https://doi.org/10.1103/PhysRevD.96.063509}{\emph{Phys. Rev. D} {\bfseries 96} (2017) 063509} [\href{https://arxiv.org/abs/1705.00552}{{\ttfamily 1705.00552}}].

\bibitem{Dimopoulos:2017tud}
K.~Dimopoulos, L.~Donaldson~Wood and C.~Owen, \emph{{Instant preheating in quintessential inflation with $\alpha$-attractors}}, \href{https://doi.org/10.1103/PhysRevD.97.063525}{\emph{Phys. Rev. D} {\bfseries 97} (2018) 063525} [\href{https://arxiv.org/abs/1712.01760}{{\ttfamily 1712.01760}}].

\bibitem{Akrami:2017cir}
Y.~Akrami, R.~Kallosh, A.~Linde and V.~Vardanyan, \emph{{Dark energy, $\alpha$-attractors, and large-scale structure surveys}}, \href{https://doi.org/10.1088/1475-7516/2018/06/041}{\emph{JCAP} {\bfseries 06} (2018) 041} [\href{https://arxiv.org/abs/1712.09693}{{\ttfamily 1712.09693}}].

\bibitem{Garcia-Garcia:2018hlc}
C.~Garc\'\i{}a-Garc\'\i{}a, E.V.~Linder, P.~Ru\'\i{}z-Lapuente and M.~Zumalac\'arregui, \emph{{Dark energy from $\alpha$-attractors: phenomenology and observational constraints}}, \href{https://doi.org/10.1088/1475-7516/2018/08/022}{\emph{JCAP} {\bfseries 08} (2018) 022} [\href{https://arxiv.org/abs/1803.00661}{{\ttfamily 1803.00661}}].

\bibitem{Herranen:2014cua}
M.~Herranen, T.~Markkanen, S.~Nurmi and A.~Rajantie, \emph{{Spacetime curvature and the Higgs stability during inflation}}, \href{https://doi.org/10.1103/PhysRevLett.113.211102}{\emph{Phys. Rev. Lett.} {\bfseries 113} (2014) 211102} [\href{https://arxiv.org/abs/1407.3141}{{\ttfamily 1407.3141}}].

\bibitem{Gialamas:2022gxv}
I.D.~Gialamas, A.~Karam and T.D.~Pappas, \emph{{Gravitational corrections to electroweak vacuum decay: metric vs. Palatini}}, \href{https://doi.org/10.1016/j.physletb.2023.137885}{\emph{Phys. Lett. B} {\bfseries 840} (2023) 137885} [\href{https://arxiv.org/abs/2212.03052}{{\ttfamily 2212.03052}}].

\bibitem{Gialamas:2023emn}
I.D.~Gialamas and H.~Veerm\"ae, \emph{{Electroweak vacuum decay in metric-affine gravity}}, \href{https://doi.org/10.1016/j.physletb.2023.138109}{\emph{Phys. Lett. B} {\bfseries 844} (2023) 138109} [\href{https://arxiv.org/abs/2305.07693}{{\ttfamily 2305.07693}}].

\bibitem{Opferkuch:2019zbd}
T.~Opferkuch, P.~Schwaller and B.A.~Stefanek, \emph{{Ricci Reheating}}, \href{https://doi.org/10.1088/1475-7516/2019/07/016}{\emph{JCAP} {\bfseries 07} (2019) 016} [\href{https://arxiv.org/abs/1905.06823}{{\ttfamily 1905.06823}}].

\bibitem{Cosme:2018nly}
C.~Cosme, J.a.G.~Rosa and O.~Bertolami, \emph{{Scale-invariant scalar field dark matter through the Higgs portal}}, \href{https://doi.org/10.1007/JHEP05(2018)129}{\emph{JHEP} {\bfseries 05} (2018) 129} [\href{https://arxiv.org/abs/1802.09434}{{\ttfamily 1802.09434}}].

\bibitem{Riotto:2002yw}
A.~Riotto, \emph{{Inflation and the theory of cosmological perturbations}}, {\emph{ICTP Lect. Notes Ser.} {\bfseries 14} (2003) 317} [\href{https://arxiv.org/abs/hep-ph/0210162}{{\ttfamily hep-ph/0210162}}].

\bibitem{Ford:2021syk}
L.H.~Ford, \emph{{Cosmological particle production: a review}}, \href{https://doi.org/10.1088/1361-6633/ac1b23}{\emph{Rept. Prog. Phys.} {\bfseries 84} (2021) } [\href{https://arxiv.org/abs/2112.02444}{{\ttfamily 2112.02444}}].

\bibitem{Felder:2001kt}
G.N.~Felder, L.~Kofman and A.D.~Linde, \emph{{Tachyonic instability and dynamics of spontaneous symmetry breaking}}, \href{https://doi.org/10.1103/PhysRevD.64.123517}{\emph{Phys. Rev. D} {\bfseries 64} (2001) 123517} [\href{https://arxiv.org/abs/hep-th/0106179}{{\ttfamily hep-th/0106179}}].

\bibitem{Kolb:2023ydq}
E.W.~Kolb and A.J.~Long, \emph{{Cosmological gravitational particle production and its implications for cosmological relics}},  \href{https://arxiv.org/abs/2312.09042}{{\ttfamily 2312.09042}}.

\bibitem{BezrukovNotebook}
F.L.~Bezrukov, ``{Standard Model $\bar{MS}$ Parameters}.'' \url{ http://www.inr.ac.ru/~fedor/SM/}.

\bibitem{Allahverdi:2020bys}
R.~Allahverdi et~al., \emph{{The First Three Seconds: a Review of Possible Expansion Histories of the Early Universe}},  \href{https://arxiv.org/abs/2006.16182}{{\ttfamily 2006.16182}}.

\bibitem{Fumagalli:2019ohr}
J.~Fumagalli, S.~Renaux-Petel and J.W.~Ronayne, \emph{{Higgs vacuum (in)stability during inflation: the dangerous relevance of de Sitter departure and Planck-suppressed operators}}, \href{https://doi.org/10.1007/JHEP02(2020)142}{\emph{JHEP} {\bfseries 02} (2020) 142} [\href{https://arxiv.org/abs/1910.13430}{{\ttfamily 1910.13430}}].

\bibitem{Enqvist:2013kaa}
K.~Enqvist, T.~Meriniemi and S.~Nurmi, \emph{{Generation of the Higgs Condensate and Its Decay after Inflation}}, \href{https://doi.org/10.1088/1475-7516/2013/10/057}{\emph{JCAP} {\bfseries 10} (2013) 057} [\href{https://arxiv.org/abs/1306.4511}{{\ttfamily 1306.4511}}].

\bibitem{Greene:2000ew}
P.B.~Greene and L.~Kofman, \emph{{On the theory of fermionic preheating}}, \href{https://doi.org/10.1103/PhysRevD.62.123516}{\emph{Phys. Rev. D} {\bfseries 62} (2000) 123516} [\href{https://arxiv.org/abs/hep-ph/0003018}{{\ttfamily hep-ph/0003018}}].

\bibitem{Rubio:2018ogq}
J.~Rubio, \emph{{Higgs inflation}}, \href{https://doi.org/10.3389/fspas.2018.00050}{\emph{Front. Astron. Space Sci.} {\bfseries 5} (2019) 50} [\href{https://arxiv.org/abs/1807.02376}{{\ttfamily 1807.02376}}].

\bibitem{Bezrukov:2012sa}
F.~Bezrukov, M.Y.~Kalmykov, B.A.~Kniehl and M.~Shaposhnikov, \emph{{Higgs Boson Mass and New Physics}}, \href{https://doi.org/10.1007/JHEP10(2012)140}{\emph{JHEP} {\bfseries 10} (2012) 140} [\href{https://arxiv.org/abs/1205.2893}{{\ttfamily 1205.2893}}].

\bibitem{Bezrukov:2007ep}
F.L.~Bezrukov and M.~Shaposhnikov, \emph{{The Standard Model Higgs boson as the inflaton}}, \href{https://doi.org/10.1016/j.physletb.2007.11.072}{\emph{Phys. Lett. B} {\bfseries 659} (2008) 703} [\href{https://arxiv.org/abs/0710.3755}{{\ttfamily 0710.3755}}].

\bibitem{Chauhan:2023pur}
G.~Chauhan and T.~Steingasser, \emph{{Gravity-improved metastability bounds for the Type-I seesaw mechanism}}, \href{https://doi.org/10.1007/JHEP09(2023)151}{\emph{JHEP} {\bfseries 09} (2023) 151} [\href{https://arxiv.org/abs/2304.08542}{{\ttfamily 2304.08542}}].

\bibitem{Kierkla:2023uzo}
M.~Kierkla, G.~Laverda, M.~Lewicki, A.~Mantziris, M.~Piani, J.~Rubio et~al., \emph{{From Hubble to Bubble}}, \href{https://doi.org/10.1007/JHEP11(2023)077}{\emph{JHEP} {\bfseries 11} (2023) 077} [\href{https://arxiv.org/abs/2309.08530}{{\ttfamily 2309.08530}}].

\bibitem{Hawking:1981fz}
S.W.~Hawking and I.G.~Moss, \emph{{Supercooled Phase Transitions in the Very Early Universe}}, \href{https://doi.org/10.1016/0370-2693(82)90946-7}{\emph{Phys. Lett. B} {\bfseries 110} (1982) 35}.

\bibitem{Coleman:1980aw}
S.R.~Coleman and F.~De~Luccia, \emph{{Gravitational Effects on and of Vacuum Decay}}, \href{https://doi.org/10.1103/PhysRevD.21.3305}{\emph{Phys. Rev. D} {\bfseries 21} (1980) 3305}.

\bibitem{Khoury:2021zao}
J.~Khoury and T.~Steingasser, \emph{{Gauge hierarchy from electroweak vacuum metastability}}, \href{https://doi.org/10.1103/PhysRevD.105.055031}{\emph{Phys. Rev. D} {\bfseries 105} (2022) 055031} [\href{https://arxiv.org/abs/2108.09315}{{\ttfamily 2108.09315}}].

\bibitem{deSalas:2015glj}
P.F.~de~Salas, M.~Lattanzi, G.~Mangano, G.~Miele, S.~Pastor and O.~Pisanti, \emph{{Bounds on very low reheating scenarios after Planck}}, \href{https://doi.org/10.1103/PhysRevD.92.123534}{\emph{Phys. Rev. D} {\bfseries 92} (2015) 123534} [\href{https://arxiv.org/abs/1511.00672}{{\ttfamily 1511.00672}}].

\bibitem{Hasegawa:2019jsa}
T.~Hasegawa, N.~Hiroshima, K.~Kohri, R.S.L.~Hansen, T.~Tram and S.~Hannestad, \emph{{MeV-scale reheating temperature and thermalization of oscillating neutrinos by radiative and hadronic decays of massive particles}}, \href{https://doi.org/10.1088/1475-7516/2019/12/012}{\emph{JCAP} {\bfseries 12} (2019) 012} [\href{https://arxiv.org/abs/1908.10189}{{\ttfamily 1908.10189}}].

\bibitem{Micha:2004bv}
R.~Micha and I.I.~Tkachev, \emph{{Turbulent thermalization}}, \href{https://doi.org/10.1103/PhysRevD.70.043538}{\emph{Phys. Rev. D} {\bfseries 70} (2004) 043538} [\href{https://arxiv.org/abs/hep-ph/0403101}{{\ttfamily hep-ph/0403101}}].

\bibitem{Micha:2002ey}
R.~Micha and I.I.~Tkachev, \emph{{Relativistic turbulence: A Long way from preheating to equilibrium}}, \href{https://doi.org/10.1103/PhysRevLett.90.121301}{\emph{Phys. Rev. Lett.} {\bfseries 90} (2003) 121301} [\href{https://arxiv.org/abs/hep-ph/0210202}{{\ttfamily hep-ph/0210202}}].

\bibitem{Micha:2003ws}
R.~Micha and I.I.~Tkachev, \emph{{Preheating and thermalization after inflation}},  in \emph{{5th Internationa Conference on Strong and Electroweak Matter}}, pp.~210--219, 2003, \href{https://doi.org/10.1142/9789812704498_0020}{DOI} [\href{https://arxiv.org/abs/hep-ph/0301249}{{\ttfamily hep-ph/0301249}}].

\bibitem{Garcia-Bellido:2008ycs}
J.~Garcia-Bellido, D.G.~Figueroa and J.~Rubio, \emph{{Preheating in the Standard Model with the Higgs-Inflaton coupled to gravity}}, \href{https://doi.org/10.1103/PhysRevD.79.063531}{\emph{Phys. Rev. D} {\bfseries 79} (2009) 063531} [\href{https://arxiv.org/abs/0812.4624}{{\ttfamily 0812.4624}}].

\bibitem{Repond:2016sol}
J.~Repond and J.~Rubio, \emph{{Combined Preheating on the lattice with applications to Higgs inflation}}, \href{https://doi.org/10.1088/1475-7516/2016/07/043}{\emph{JCAP} {\bfseries 07} (2016) 043} [\href{https://arxiv.org/abs/1604.08238}{{\ttfamily 1604.08238}}].

\bibitem{Planck:2018jri}
{\scshape Planck} collaboration, \emph{{Planck 2018 results. X. Constraints on inflation}}, \href{https://doi.org/10.1051/0004-6361/201833887}{\emph{Astron. Astrophys.} {\bfseries 641} (2020) A10} [\href{https://arxiv.org/abs/1807.06211}{{\ttfamily 1807.06211}}].

\bibitem{Shaposhnikov:1987tw}
M.E.~Shaposhnikov, \emph{{Baryon Asymmetry of the Universe in Standard Electroweak Theory}}, \href{https://doi.org/10.1016/0550-3213(87)90127-1}{\emph{Nucl. Phys. B} {\bfseries 287} (1987) 757}.

\bibitem{Wagner:2023vqw}
C.E.M.~Wagner, \emph{{Electroweak Baryogenesis and Higgs Physics}}, \href{https://doi.org/10.31526/lhep.2023.466}{\emph{LHEP} {\bfseries 2023} (2023) 466} [\href{https://arxiv.org/abs/2311.06949}{{\ttfamily 2311.06949}}].

\bibitem{Gouttenoire:2021jhk}
Y.~Gouttenoire, G.~Servant and P.~Simakachorn, \emph{{Kination cosmology from scalar fields and gravitational-wave signatures}},  \href{https://arxiv.org/abs/2111.01150}{{\ttfamily 2111.01150}}.

\bibitem{Punturo:2010zz}
M.~Punturo et~al., \emph{{The Einstein Telescope: A third-generation gravitational wave observatory}}, \href{https://doi.org/10.1088/0264-9381/27/19/194002}{\emph{Class. Quant. Grav.} {\bfseries 27} (2010) 194002}.

\bibitem{Harry:2006fi}
G.M.~Harry, P.~Fritschel, D.A.~Shaddock, W.~Folkner and E.S.~Phinney, \emph{{Laser interferometry for the big bang observer}}, \href{https://doi.org/10.1088/0264-9381/23/15/008}{\emph{Class. Quant. Grav.} {\bfseries 23} (2006) 4887}.

\bibitem{LISA}
P.~{Amaro-Seoane} et~al., \emph{{Laser Interferometer Space Antenna}},  \href{https://arxiv.org/abs/1702.00786}{{\ttfamily 1702.00786}}.

\bibitem{Garcia-Bellido:1997hex}
J.~Garcia-Bellido and A.D.~Linde, \emph{{Preheating in hybrid inflation}}, \href{https://doi.org/10.1103/PhysRevD.57.6075}{\emph{Phys. Rev. D} {\bfseries 57} (1998) 6075} [\href{https://arxiv.org/abs/hep-ph/9711360}{{\ttfamily hep-ph/9711360}}].

\bibitem{Garcia-Bellido:2001dqy}
J.~Garcia-Bellido and E.~Ruiz~Morales, \emph{{Particle production from symmetry breaking after inflation}}, \href{https://doi.org/10.1016/S0370-2693(02)01820-8}{\emph{Phys. Lett. B} {\bfseries 536} (2002) 193} [\href{https://arxiv.org/abs/hep-ph/0109230}{{\ttfamily hep-ph/0109230}}].

\bibitem{Greene:1997fu}
P.B.~Greene, L.~Kofman, A.D.~Linde and A.A.~Starobinsky, \emph{{Structure of resonance in preheating after inflation}}, \href{https://doi.org/10.1103/PhysRevD.56.6175}{\emph{Phys. Rev. D} {\bfseries 56} (1997) 6175} [\href{https://arxiv.org/abs/hep-ph/9705347}{{\ttfamily hep-ph/9705347}}].

\bibitem{Fan:2021otj}
J.~Fan, K.D.~Lozanov and Q.~Lu, \emph{{Spillway Preheating}}, \href{https://doi.org/10.1007/JHEP05(2021)069}{\emph{JHEP} {\bfseries 05} (2021) 069} [\href{https://arxiv.org/abs/2101.11008}{{\ttfamily 2101.11008}}].

\bibitem{Kofman:1997yn}
L.~Kofman, A.D.~Linde and A.A.~Starobinsky, \emph{{Towards the theory of reheating after inflation}}, \href{https://doi.org/10.1103/PhysRevD.56.3258}{\emph{Phys. Rev. D} {\bfseries 56} (1997) 3258} [\href{https://arxiv.org/abs/hep-ph/9704452}{{\ttfamily hep-ph/9704452}}].

\bibitem{Enqvist:2015sua}
K.~Enqvist, S.~Nurmi, S.~Rusak and D.~Weir, \emph{{Lattice Calculation of the Decay of Primordial Higgs Condensate}}, \href{https://doi.org/10.1088/1475-7516/2016/02/057}{\emph{JCAP} {\bfseries 02} (2016) 057} [\href{https://arxiv.org/abs/1506.06895}{{\ttfamily 1506.06895}}].

\bibitem{Bodeker:2007fw}
D.~Bodeker and K.~Rummukainen, \emph{{Non-abelian plasma instabilities for strong anisotropy}}, \href{https://doi.org/10.1088/1126-6708/2007/07/022}{\emph{JHEP} {\bfseries 07} (2007) 022} [\href{https://arxiv.org/abs/0705.0180}{{\ttfamily 0705.0180}}].

\end{thebibliography}\endgroup

\end{document}